\documentclass[11pt,a4paper]{article}
\usepackage{jcappub,graphicx,subfigure}
\usepackage{epsfig,multicol,bbm,latexsym}

\title{Constraints on the dark matter annihilation from Fermi-LAT
observation of M31}

\author{Zhengwei Li$^{a,b}$, Xiaoyuan Huang$^c$, Qiang Yuan$^{a}$\footnote{For 
correspondence.}, and Yupeng Xu$^b$}

\affiliation{
$^a$Key Laboratory of Dark Matter and Space Astronomy, Purple Mountain 
Observatory, Chinese Academy of Sciences, Nanjing 210008, P.R.China\\
$^b$Key Laboratory of Particle Astrophysics, Institute of High Energy 
Physics, Chinese Academy of Sciences, Beijing 100049, P.R. China\\
$^c$Physik-Department T30d, Technische Universit\"at M\"unchen, 
James-Franck-Stra\ss{}e, D-85748 Garching, Germany}
\emailAdd{lizw@ihep.ac.cn}
\emailAdd{huangxiaoyuan@gmail.com}
\emailAdd{yuanq@pmo.ac.cn}
\emailAdd{xuyp@ihep.ac.cn}

\abstract{Gamma-ray is a good probe of dark matter (DM) particles in 
the Universe. We search for the DM annihilation signals in the direction 
of the Andromeda galaxy (M31) using 7.5 year Fermi-LAT pass 8 data. 
Similar to Pshirkov et al. (2016), we find that there is residual
excess emission from the direction of M31 if only the galactic disk
as traced by the far infrared emission is considered. Adding a point-like 
source will improve the fitting effectively, although additional slight 
improvements can be found if an extended component such as a uniform
disk or two bubbles is added instead. Taking the far infrared disk plus a 
point source as the background model, we search for the DM annihilation 
signals in the data. We find that there is strong degeneracy between the 
emission from the galaxy and that from 10s GeV mass DM annihilation in
the main halo with quark final state. However, the required DM 
annihilation cross section is about $10^{-25}-10^{-24}$ cm$^3$s$^{-1}$, 
orders of magnitude larger than the constraints from observations of dwarf 
spheroidal galaxies, indicating a non-DM origin of the emission. 
If DM subhalos are taken into account, the degeneracy is broken.
When considering the enhancement from DM subhalos, the constraints 
on DM model parameters are comparable to (or slightly weaker than) those 
from the population of dwarf spheroidal galaxies. We also discuss the 
inverse Compton scattering component from DM annihilation induced 
electrons/positrons. For the first time we include an energy dependent 
template of the inverse Compton emission (i.e., a template cube) in the 
data analysis to take into account the effect of diffusion of charged 
particles. We find a significant improvement of the constraints in the high 
mass range of DM particles after considering the inverse Compton emission.}

\keywords{dark matter, gamma ray, M31}

\arxivnumber{1312.7609}

\begin{document}
\maketitle
\flushbottom

\section{Introduction}
\label{sec:intro}

It has been over 80 years since the first discovery of dark matter (DM)
by F. Zwicky in the 1930s~\cite{1933AcHPhZwicky}, but the particle nature
of DM remains one of the biggest unsolved problems of physics. Many
astronomical observations show that the majority of DM should be made of
non-baryonic, non-luminous, and cold matter. The most popular candidate 
is a weakly interacting massive particle (WIMP), which could just
produce the right relic density of DM assuming thermal freezing out of
DM particles in the early Universe. Widely discussed candidates include
the lightest particle in the supersymmetric extension of the standard
model (SM) or the universal extra dimention theory \cite{1996PhRJungman,
2005PhRBertone,2007PhRHooper}. Self-annihilation or decay of WIMPs into
SM particles can produce charged particles and $\gamma$-rays with signatures
different from the ordinary astrophysical background, which may be detectable
in the cosmic ray or multi-wavelength electromagnetic observations
\cite{2008NaturChang,2009NaturAdriani,2006A&AColafrancesco}.

The Fermi Large Area Telescope (Fermi-LAT)~\cite{2009ApJAtwood}, launched 
in 2008, is up to now the most sensitive detector for GeV $\gamma$-rays.
It significantly improves the sensitivity of the searching for DM
particles in space. Many targets have been studied thoroughly to search 
for DM signals based on the Fermi-LAT data, including the Milky Way
dwarf galaxies \cite{Abdo:2010ex,2011PhRvLAckermann,2011PhRvLGeringer-Sameth,
2013JCAPTsai,2013Ackermann,Ackermann:2015zua,Geringer-Sameth:2014qqa},
clusters of galaxies \cite{2010JCAPDugger,2010PhRvDYuan,2010JCAPAckermann,
2012aJCAPHuang,2012PhRvDCombet,Han:2012uw}, the center and halo of the 
Milky Way \cite{2011JCAPEllis,2011PhLBHooper,2011PhRvDHooper,
2012PhRvDAbazajian,2012ApJAckermann,2012bJCAPHuang,2013APhHooper}, 
the globular clusters \cite{2012JCAPFeng} and so on. Also there were 
efforts to search for the monochromatic line emission from the Fermi-LAT 
data \cite{2010PhRvLAbdo,2012JCAPBringmann,2012JCAPWeniger,
2013PhRvDAckermann,Ackermann:2015lka,Liang:2016pvm}. Some tentative 
candidates of DM signals were reported from the $\gamma$-ray data of the 
Galactic center region \cite{2011PhLBHooper,2011PhRvDHooper} and a few 
ultra-faint dwarf galaxies \cite{Geringer-Sameth:2015lua,Li:2015kag}. 
However, no consistent and conclusive evidence of DM signals can be 
established yet. Effective constraints on the DM model parameters can 
be set according to the non-detection of DM signals.

As the nearest (with a distance of $785\pm25$ kpc \cite{2010A&ACorbelli}) 
large galaxy, the Andromeda galaxy (M31) is also a potentially good target 
for DM searches \cite{2004APhFalvard,2010JCAPDugger,2012JCAPWatson,
2013PhRvDEgorov}. The location of M31 is away from the Galactic plane 
($b\approx-22^{\circ}$) which will be less polluted by the strong 
Galactic foreground emission. Although it is relatively faint, M31 
has been detected at $\gamma$-ray band by the Fermi-LAT 
\cite{2010A&AAbdo,Bird:2015npa,2016MNRAS.459L..76P}. Its $\gamma$-ray 
spectrum and luminosity are consistent with predictions from cosmic ray 
collisions with the interstellar medium (ISM), just as those established 
for the Milky Way. Taking the detected flux as the upper limit on the 
DM induced $\gamma$-ray emission from M31, conservative limits on the 
DM annihilation cross section or decay lifetime were derived
\cite{2010A&AAbdo,2010JCAPDugger}.

In this work, we use the 7.5 year data of the Fermi-LAT to improve the
sensitivity of searches for DM annihilation signal in M31. The prediction 
of DM annihilation induced $\gamma$-ray flux depends on the detailed 
structure of the DM distribution. Currently the knowledge about the DM 
distribution, especially at small scales, largely relies on numerical 
simulations \cite{1996ApJNavarro,1997ApJNavarro,2008NaturDiemand,
2008NaturSpringel}. High-resolution numerical simulations show that a 
large population of subhalos which could extend down to very low 
masses exist in the main DM halo \cite{2008NaturDiemand,2008NaturSpringel,
2008MNRASSpringel}. It was pointed out that subhalos will alter the 
spatial extension of the DM induced $\gamma$-ray signals and affect
the search strategy of DM particles \cite{2012MNRASGao}. Therefore we will
discuss different spatial distributions of the M31 halo as well as the
subhalo population in this work. The spatial templates of DM
annihilation induced $\gamma$-rays are built and implemented in the
likelihood analysis of the Fermi-LAT data. Typical annihilation channels
to a pair of $b$ quarks, gauge bosons, or charged leptons will be
discussed. We will also discuss the inverse Compton scattering (ICS) 
component from DM annihilation induced electrons/positrons. And for the 
first time we will include an energy dependent template of the ICS emission 
(i.e., a template cube) in the data analysis to take the effect of 
diffusion of charged particles into account.

\section{Gamma ray flux from DM annihilation in M31}
\label{sec:2}

Assuming the DM particles are Majorana fermions, the $\gamma$-ray flux 
from DM annihilation as a function of energy $E$ and direction $\theta$ 
can be written as
\begin{equation}
\label{eq:x}
\frac{d\Phi}{dEd\Omega}(\theta,E) = W(E)\times J(\theta) 
=\frac{\langle\sigma v\rangle}{2m^{2}_{\chi}}\frac{dN_{\gamma}}{dE} 
\times \frac{1}{4\pi}\int_{l.o.s}ds\,\rho^{2}_{\rm DM}(r(s)).
\end{equation}
where the integral is computed along the line of sight (l.o.s.), $W(E)$ 
and $J(\theta)$ represent the energy and spatial dependent parts, $m_{\chi}$ 
is the mass of the DM particle, $\langle\sigma v\rangle$ is the velocity 
weighted average pair annihilation cross section, $dN_{\gamma}/dE$ is the 
energy spectrum of $\gamma$-rays for one annihilation which is computed 
with the Pythia simulation code \cite{2006JHEPSjstrand}, and 
$\rho_{\rm DM}(r)$ denotes the DM density distribution.

\subsection{DM density profile}

The expected $\gamma$-ray signal depends crucially on the DM density
profile $\rho_{\rm DM}(r)$ of the halo. The universal density profile
suggested by Navarro, Frenk and White (hereafter NFW) found in numerical 
simulations \cite{1997ApJNavarro} is widely adopted to describe the DM 
density distribution in the main galactic halo
\begin{equation}
\label{eq:nfw}
\rho_{\rm NFW}(r)=\frac{\rho_{s}}{(r/r_{s})(1+r/r_{s})^{2}},
\end{equation}
where $r$ is the radial distance from the halo center, $\rho_{s}$ and
$r_{s}$ are the density normalization and scale radius respectively.

With even higher resolution of numerical simulations, the asymptotic flat 
Einasto (EIN) profile \cite{1965TrAlmEinasto} is shown to better fit the
simulation results at the central part \cite{2004MNRASNavarro,
2010MNRASNavarro}, which reads
\begin{equation}
\label{eq:einasto}
\rho_{\rm EIN}(r)=\rho_{s}\cdot\exp\left[-\frac{2}{\alpha}\left(\left(
\frac{r}{r_{s}}\right)^{\alpha}-1\right)\right].
\end{equation}

We should keep in mind that the density profile may have uncertainties
if we extrapolate the simulation results down to very small scales.
Furthermore, the above profiles were derived based on the pure DM 
simulations. The DM density profile may get changed when the effect of 
baryons are taken into account \cite{2010MNRASDuffy}. Therefore we also 
discuss the isothermal (ISO) density profile as a conservative example.
The isothermal density profile is given by \cite{1980ApJSBahcall}
\begin{equation}
\label{eq:isothermal}
\rho_{\rm ISO}(r)=\frac{\rho_{s}}{1+(r/r_{s})^{2}}.
\end{equation}

The measured virial mass of M31 ranges from $0.7\times10^{12}$ to 
$2.1\times10^{12}$ M$_{\odot}$ \cite{2013MNRASFardal,2000ApJEvans,
2003ApJWidrow,2005ApJIbata,2008MNRASSeigar,2010MNRASWatkins}.
In this work we assume the virial mass of M31 is $M_{\rm vir}\approx1.0
\times10^{12}$ M$_{\odot}$ with a virial radius of $r_{\rm vir}\approx205$ kpc.
These values are similar with that of our Milky Way \cite{2008ApJXue}.
Therefore we adopt similar profile parameters for M31 as that of the
Milky Way halo. The parameters are shown in Table \ref{tab:profile}
\cite{2009JCAPBertone,1996ApJNavarro,2006AJMerritt}. Note that the
normalization $\rho_s$ is derived through normalizing the total mass to
$M_{\rm vir}$.

\begin{table}[htb]
\centering
\begin{tabular}{ccc}
\hline
\hline
halo model & $r_{s}$ (kpc) & $\rho_{s}$ (GeV/cm$^{3}$)\\
\hline
NFW & 20 &  0.26\\
EIN & 20 & 0.06\\
ISO & 5 & 0.46\\
\hline
\hline
\end{tabular}
\caption{\label{tab:profile}Parameters of different density profiles of M31.}
\end{table}

\begin{figure}[htb]
\centering
\hspace{-15mm}\includegraphics[width=0.53\textwidth]{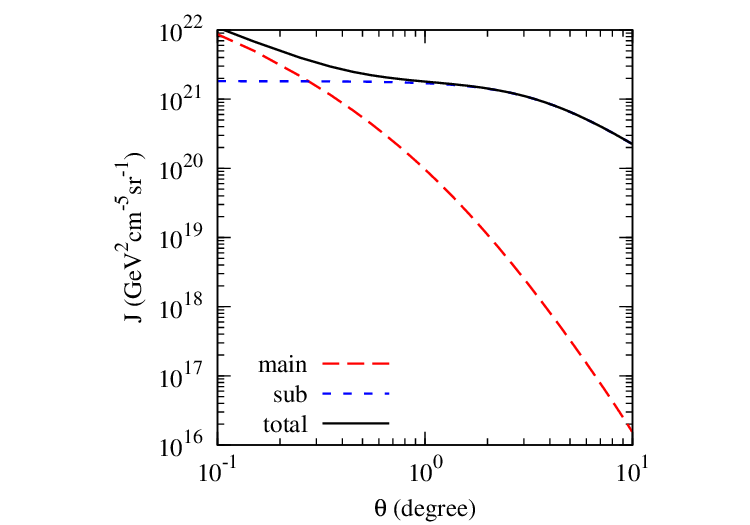}
\hspace{-5mm}\includegraphics[width=0.4\textwidth]{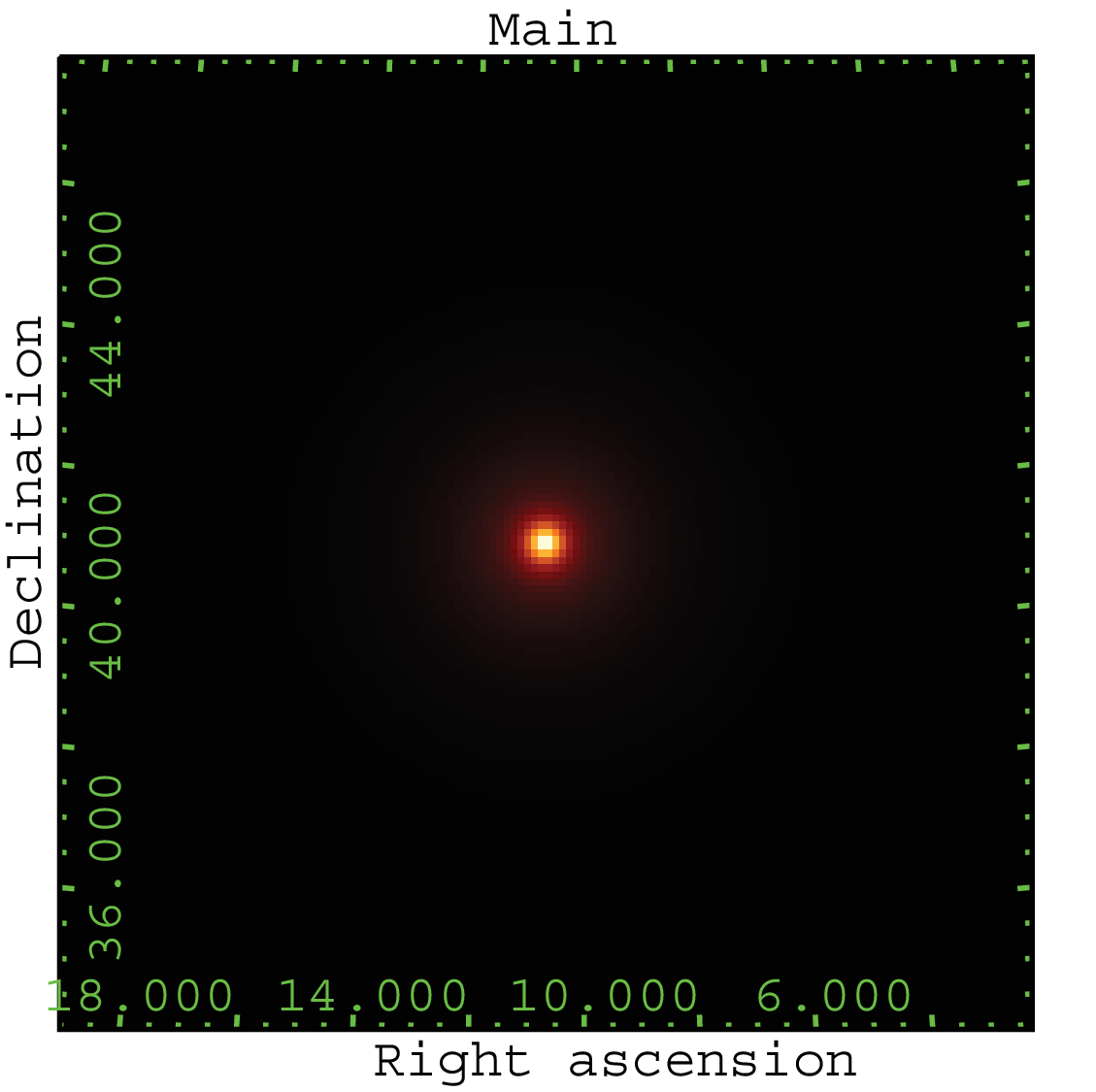}
\includegraphics[width=0.4\textwidth]{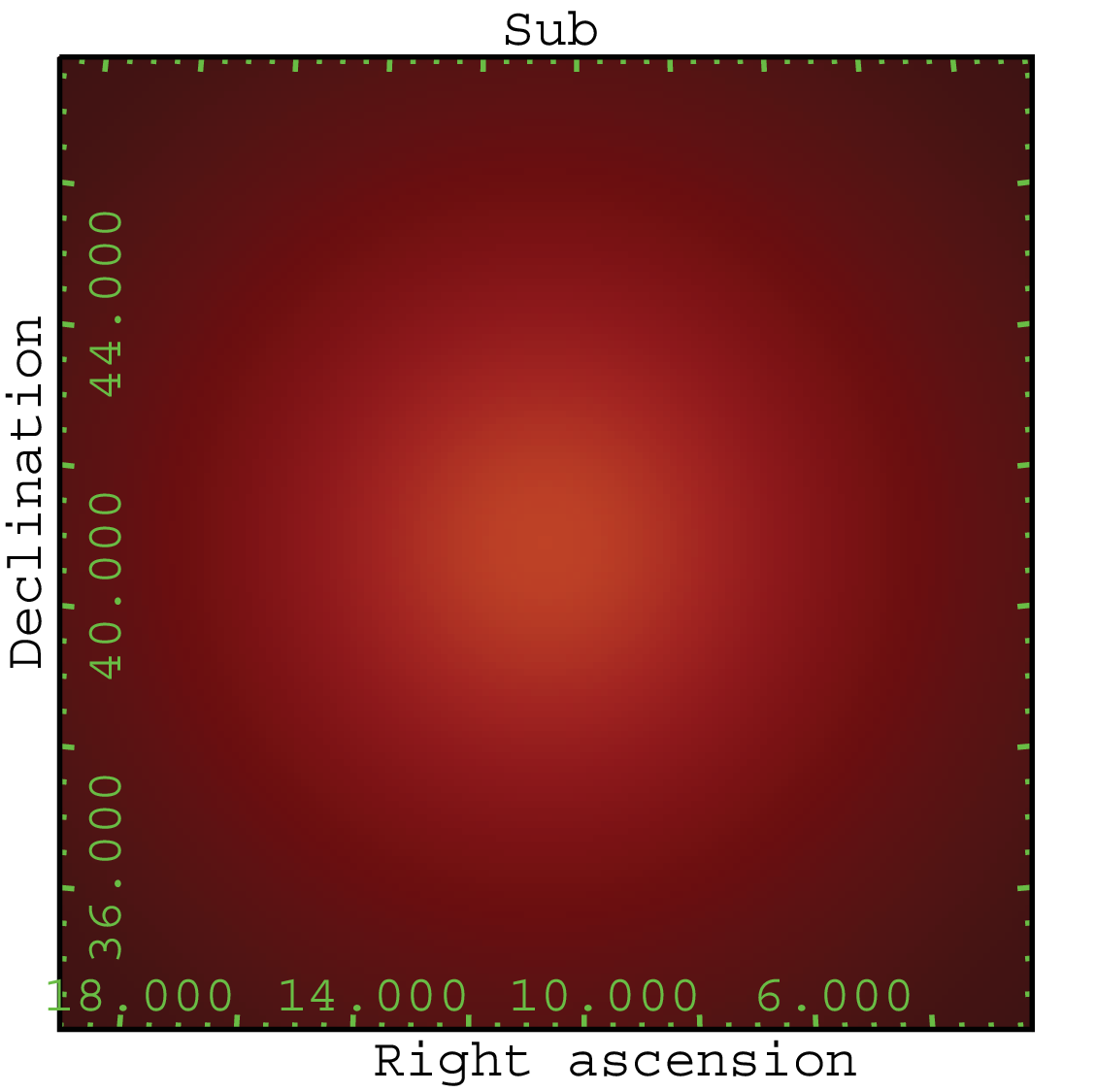}
\includegraphics[width=0.4\textwidth]{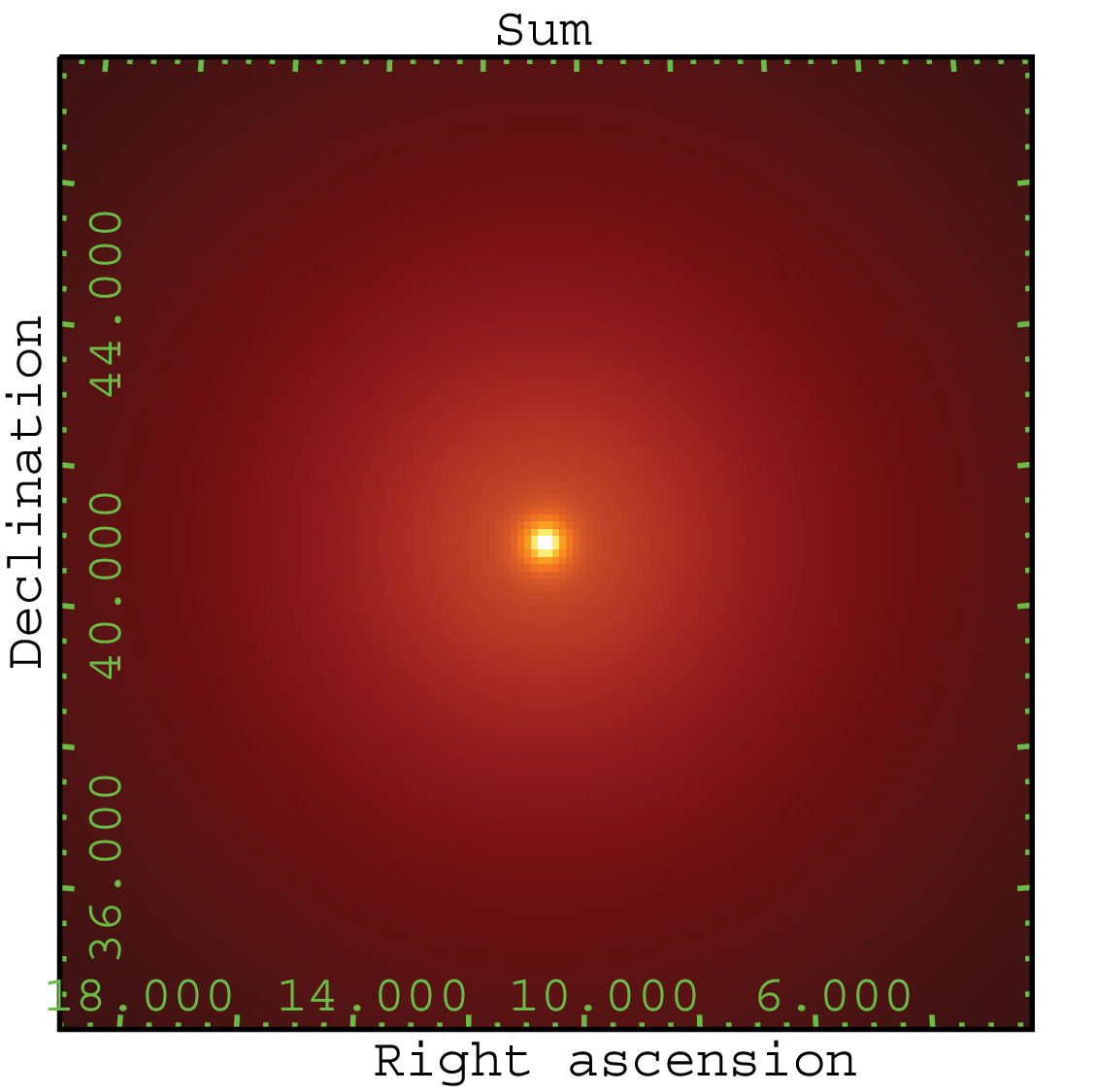}
\caption{The astrophysical factor $J$ of the DM annihilation in the main
halo and subhalos. The NFW profile of the main halo is assumed. The top-left
panel shows the 1-dimensional distribution, and the others are 2-dimensional
skymaps in a $14^{\circ}\times14^{\circ}$ region centered on M31 of the 
main halo (top-right), subhalos (bottom-left) and the sum (bottom-right).
\label{fig:Jf}}
\end{figure}

\subsection{Substructures}

Abundant subhalos are expected to form based on the structure formation
scenario of cold DM particles. Since the DM annihilation rate depends on 
the density square of DM, the existence of subhalos can boost the
annihilation signal considerably with respect to the signal from the main
halo. Also the spatial distribution of subhalos differ significantly from
that of the main halo \cite{2012MNRASGao}. Here we adopt the extracted 
subhalo distribution from the high resolution simulation Phoenix 
\cite{2012MNRASGao}. Extrapolating the mass of subhalos to a minimum 
mass of $\sim10^{-6}M_{\odot}$, the boost factor of the DM annihilation 
from subhalos is \cite{2012MNRASGao}
\begin{equation}
\label{eq:b}
b(M_{\rm vir})=L_{\rm sub}/L_{\rm main}=1.6\times 10^{-3}
(M_{\rm vir}/M_{\odot})^{0.39},
\end{equation}
which is about 76.6 for a virial mass of $M_{\rm vir}=10^{12}$ M$_{\odot}$.
The angular distribution of the annihilation $J$-factor of subhalos 
within the virial radius is
\begin{eqnarray}
\label{eq:sub}
J_{\rm sub}(\theta)&=&\frac{b(M_{\rm vir})L_{\rm main}}{4\pi d^2} 
\times f(\theta) \nonumber\\
&=&\frac{b(M_{\rm vir})L_{\rm main}}{4\pi d^2} \times \frac{16d^2}
{\pi\ln17\,r_{\rm vir}^2}\frac{1}{1+(4d\sin\theta/r_{\rm vir})^2},
\end{eqnarray}
where $d$ is the distance of the halo center to the Earth, $L_{\rm main}=
\int_{\rm main}\rho_{\rm DM}^{2}dV$ is the total annihilation luminosity 
of the main halo. 

Here we assume a ``smooth'' distribution of subhalos in the 
main halo. Such an approximation is acceptable considering the finite 
spatial resolution of $\gamma$-ray detectors. Considering the maximum 
subhalo with a mass of $\sim10^{10}$ M$_{\odot}$, its virial radius is 
about 40 kpc. For an NFW density profile, most of the DM annihilation 
occurs within a few percent of the virial radius \cite{2008NaturSpringel}, 
which corresponds to an angular radius of $\lesssim 0^{\circ}.1$ for a 
distance of $\sim 800$ kpc. Given the best resolution angle of the 
Fermi-LAT is about $0^{\circ}.1$ \cite{2009ApJAtwood} and most of the 
photons from the direction of M31 have energies below $\sim 10$ GeV, 
even the brightest subhalo in M31 is nearly unresolvable by the Fermi-LAT. 
Therefore it is reasonable to assume a ``smooth'' distribution of all 
subhalos. 

The total $J$-factor is thus the sum of the two components: $J_{\rm tot}
(\theta)=J_{\rm main}(\theta)+J_{\rm sub}(\theta)$. Figure \ref{fig:Jf} 
shows the 1-dimensional (top-left panel) and 2-dimensional distributions 
of $J_{\rm main}$ (top-right), $J_{\rm sub}$ (bottom-left) and $J_{\rm tot}$ 
(bottom-right). The results show that the $\gamma$-ray flux from the main 
halo is highly concentrated, while the subhalo contribution is much flatter. 
These sky distributions will be used as spatial templates for the analysis 
of the Fermi-LAT data in the following.

\section{Data analysis}

The detection of $\gamma$-ray emission from M31 was reported in Ref.
\cite{2010A&AAbdo}. The $\gamma$-ray emission slightly favors (at $1.8\sigma$ 
confidence level) a spatially extended source coincident with the 100 $\mu$m 
far infrared image of the Improved Reprocessing of the IRAS Survey (IRIS) 
\cite{2005ApJSMivilleDeschenes}. The best fitting spectral index is 
$\Gamma\approx 2.1$ for an extended template, and $\Gamma\approx2.5$ for a 
point source assumption, respectively. Slightly different values were 
reported in Refs. \cite{Bird:2015npa,2016MNRAS.459L..76P}, probably due 
to different energy cuts of those analyses.

The data used in our analysis are the Pass 8 events with ``SOURCE'' event
class of the Fermi-LAT data\footnote{http://fermi.gsfc.nasa.gov/ssc/data}
recorded between 4 Aug 2008 and 1 Feb 2016. We select the events with 
energies between 200 MeV and 500 GeV, and apply the zenith angle cut 
$\theta<100^{\circ}$ to suppress the contribution from the Earth limb. 
We further select events when the satellite's rocking angle is less than 
$52^{\circ}$. The radius of the region-of-interest (ROI) is taken as
$10^{\circ}$ around M31. The selected events are binned with $0.1^{\circ}
\times 0.1^{\circ}$ spatial pixels and 30 logarithmic energy bins. 
We employ the binned likelihood analysis method to analyze the data with 
the LAT Scientific Tools {\tt v10r0p5}. The instrument response function 
(IRF) adopted is {\tt P8R2\_SOURCE\_V6}. For the diffuse backgrounds we 
use the Galactic diffuse model {\tt gll\_iem\_v06.fits} and the isotropic 
background spectrum {\tt iso\_P8R2\_SOURCE\_V6\_v06.txt} as recommended 
by the Fermi-LAT collaboration\footnote
{http://fermi.gsfc.nasa.gov/ssc/data/access/lat/BackgroundModels.html}.

\begin{figure}
\centering
\includegraphics[width=.45\textwidth]{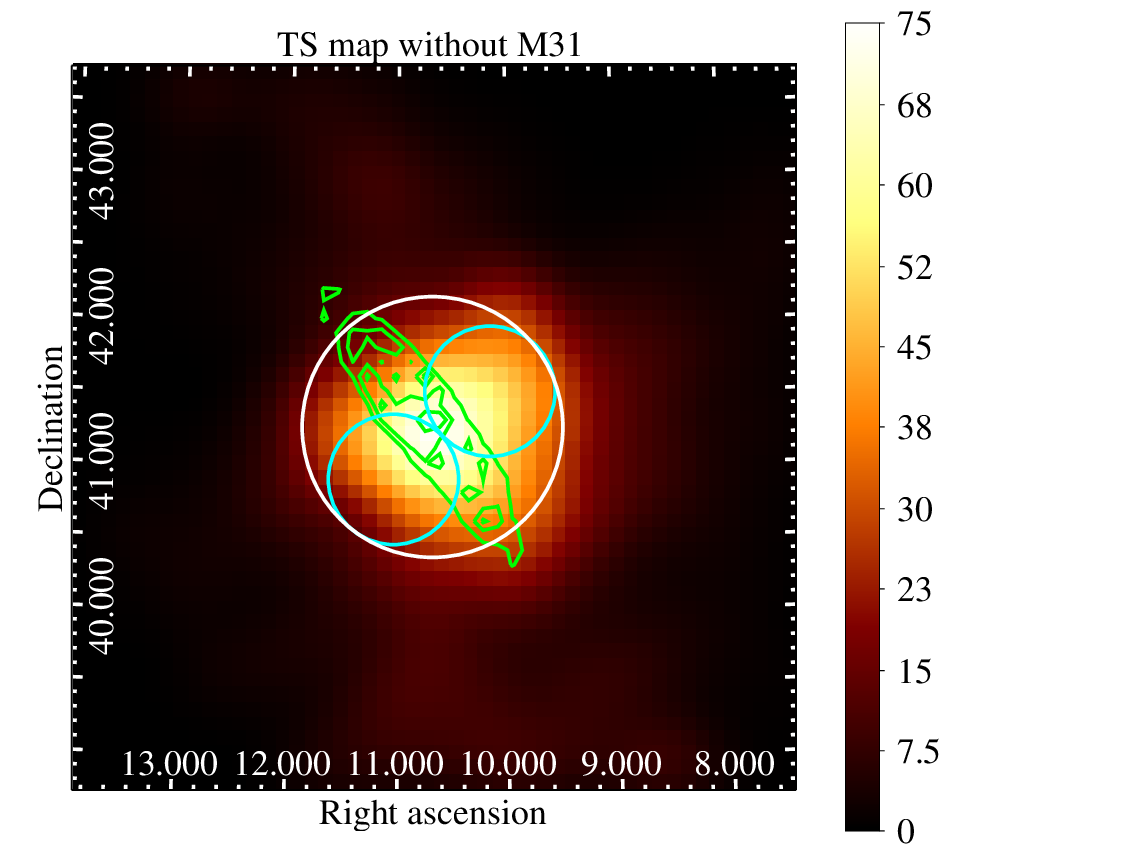}
\includegraphics[width=.45\textwidth]{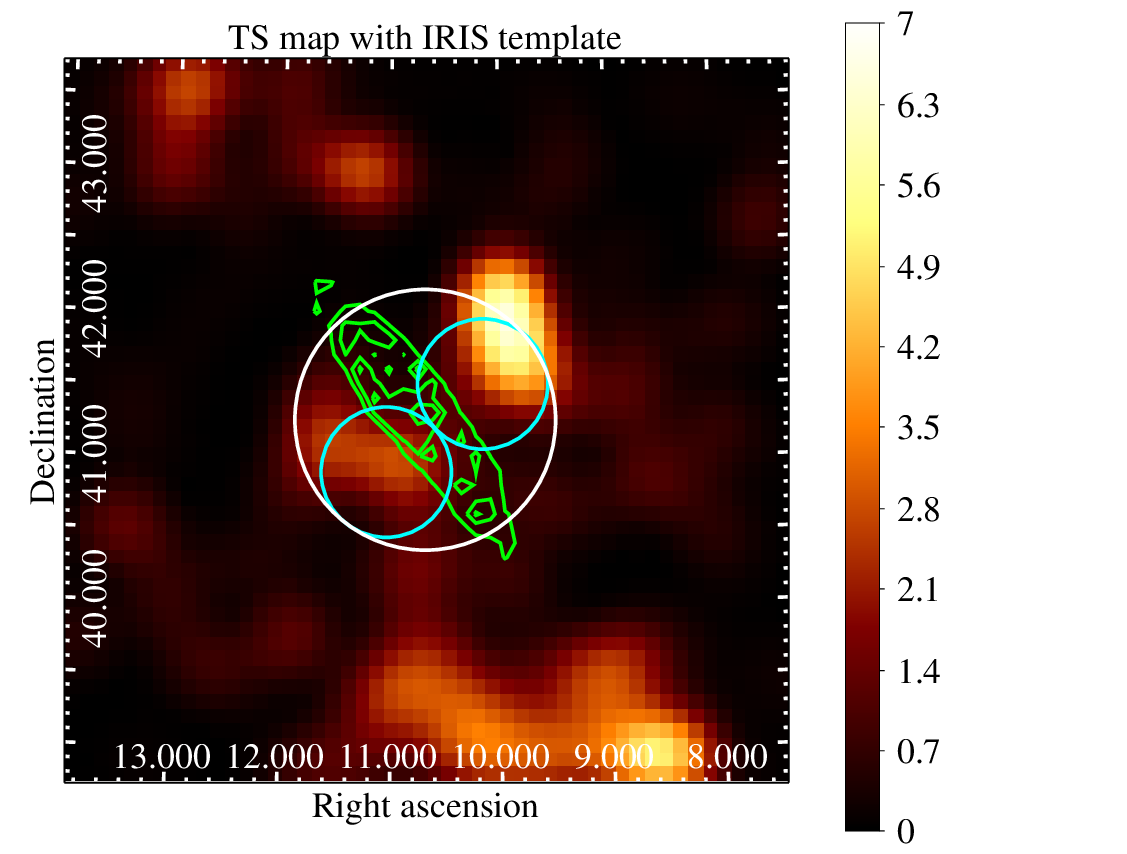}
\includegraphics[width=.45\textwidth]{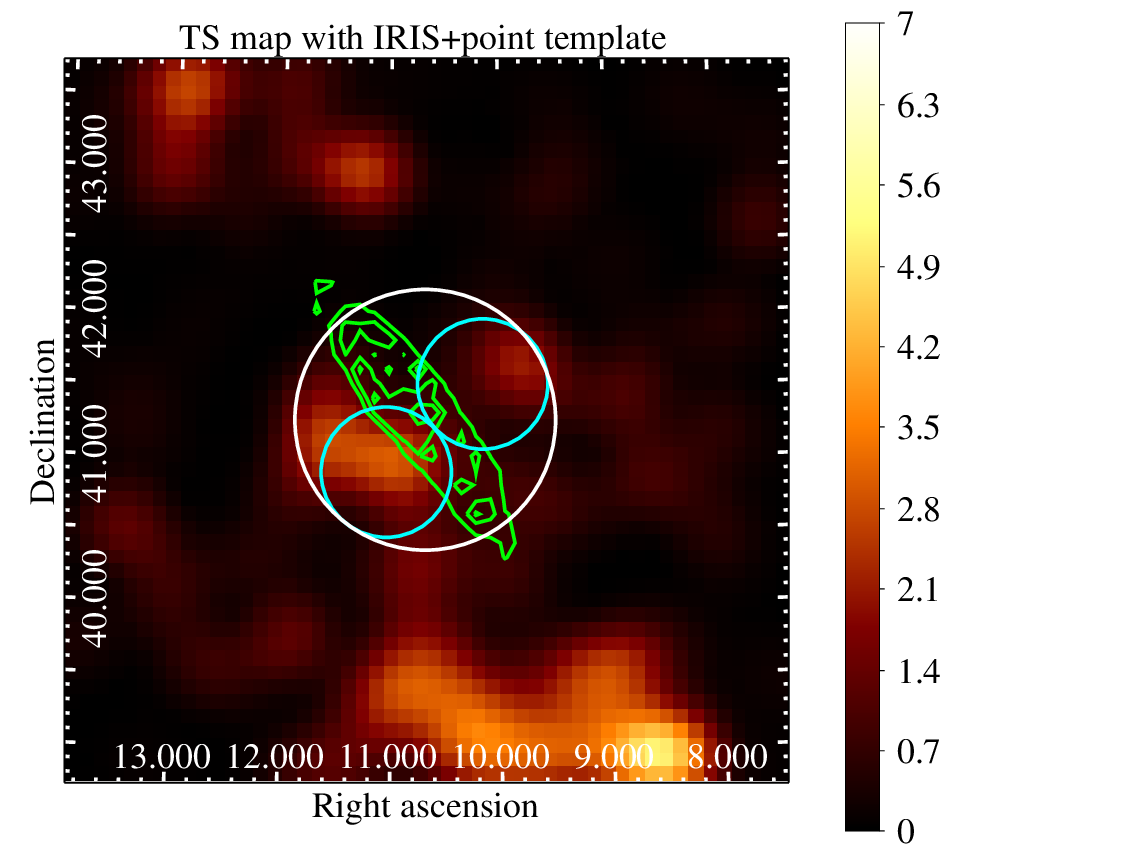}
\includegraphics[width=.45\textwidth]{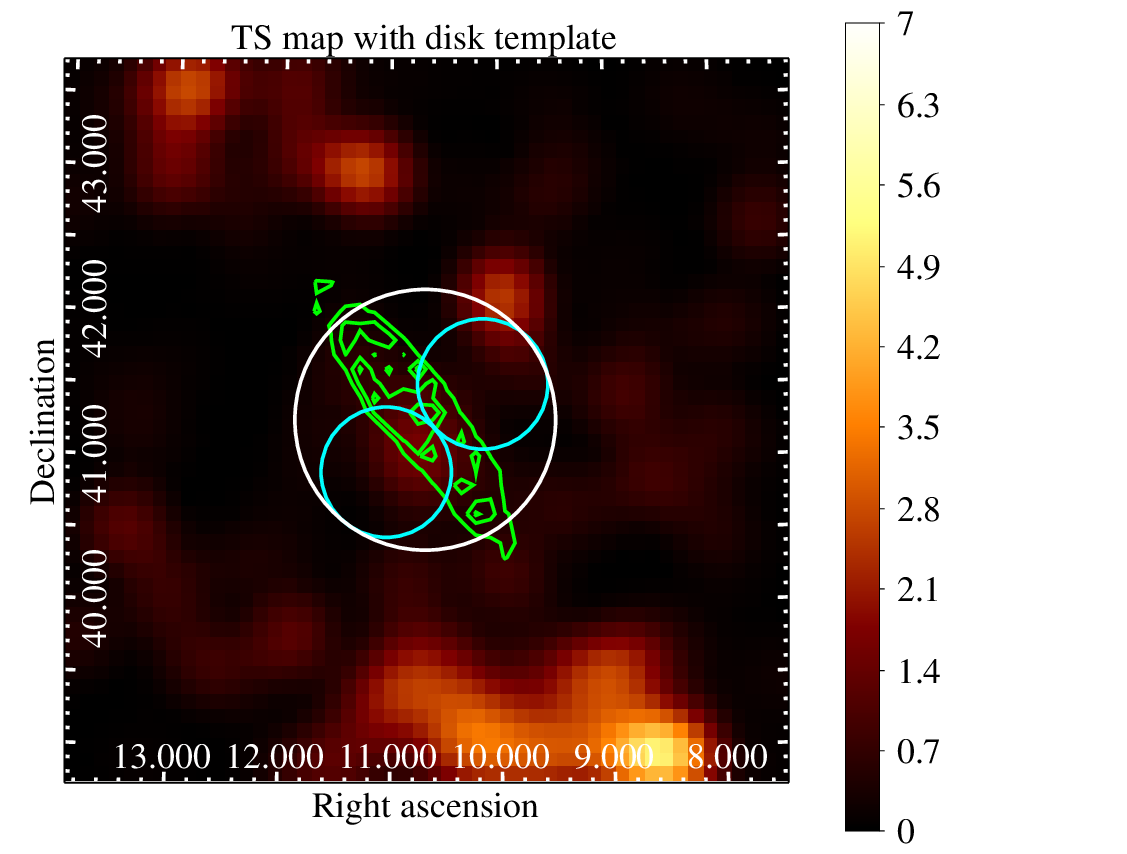}
\caption{TS maps of $5^{\circ}\times 5^{\circ}$ region centered on M31.
Top-left: excluding M31 from the model; top-right: including IRIS template
of M31; bottom-left: including IRIS template and an additional point source; 
bottom-right: including a $0^{\circ}.9$ radius uniform disk template.
Overlaid are the IRIS 100 $\mu$m contours of M31 (green), a $0^{\circ}.9$ 
radius circle (white), and two $0^{\circ}.45$ radius circles (cyan) 
representing the bubbles adopted in Ref. \cite{2016MNRAS.459L..76P}. 
The TS maps are smoothed with Gaussian kernels with $\sigma=0^{\circ}.2$.}
\label{fig:ts}
\end{figure}

We use the likelihood tool {\tt gtlike} to perform the analysis. 
The source model XML file is generated using the user contributed 
{\tt make3FGLxml.py}
tool\footnote{http://fermi.gsfc.nasa.gov/ssc/data/analysis/user/}
based on the 3FGL source catalog~\cite{Acero:2015hja}. The spatial
template of M31 is also adopted to be the IRIS 100 $\mu$m far infrared
image. The spectrum of M31 is modeled as a power-law function.  
Setting all the source parameters within the ROI and the normalizations 
of the two diffuse backgrounds free, we do the global fit to the data. 
The {\it Test Statistic}\footnote{Defined as $-2\ln({\mathcal L}/{\mathcal 
L}_0)$, where ${\mathcal L}$ (${\mathcal L}_0$) is the likelihood of
the model with (without) the target source \cite{1996ApJMattox}.} (TS)
value of the IRIS template is 91.5, and the fitting spectral index is 
$2.3\pm0.1$. We also test the point source assumption, and get TS $=59.9$
and $\Gamma=2.5\pm0.1$. The spectral index for point source assumption is
softer than that of the extended source assumption, which is consistent 
with Refs. \cite{2010A&AAbdo,Bird:2015npa,2016MNRAS.459L..76P}.

To check that whether the current model describes the data well or not,
we generate the TS maps of the $5^{\circ}\times5^{\circ}$ region centered 
on M31. The top-left panel of Figure \ref{fig:ts} shows the TS map 
without M31 in the model, overlaid with the IRIS 100 $\mu$m image contours 
(green; \cite{2005ApJSMivilleDeschenes}). Strong $\gamma$-ray emission at 
the location of M31 can be seen in the TS map. The top-right panel shows 
the TS map with the IRIS template in the model. A point-like excess at 
($00^h39^m.8$, $41^{\circ}52'$) is seen. We note that the location of this 
source deviates from that of the satellite galaxy M110 by about 
$0^{\circ}.22$, which suggests a non-M110 origin of it. We then add a new 
point source at this position with a power-law spectrum and re-do the fit. 
The overall $-\ln{\mathcal L}$ value decreases by about 8, and the TS value 
of this new point source is found to be about 16 which has some degeneracy w
ith the extended emission of M31. The residual TS map for such a model 
is shown by the bottom-left panel of Figure \ref{fig:ts}. No significant 
excess can be seen from this residual TS map. 

In Ref. \cite{2016MNRAS.459L..76P} it was claimed that halo-like (or more
specifically, bubble-like) excesses exist in the direction of M31. We test
this result in our analysis. We first adopt a $0^{\circ}.9$ radius uniform 
disk \cite{2016MNRAS.459L..76P} centered on M31 as the spatial template, 
and find that the goodness-of-fit ($-\ln{\mathcal L}$) is comparable to 
that of the IRIS + point case. The residual TS map is also similar to that
of the IRIS + point case, as shown by the bottom-right panel of Figure 
\ref{fig:ts}. Then we test the model with both the IRIS template and the 
$0^{\circ}.9$ uniform disk (two $0^{\circ}.45$ bubbles), and find a slight 
change of the overall goodness-of-fit (see Table \ref{tab:fit} for a 
summary of the fitting results). Due to the limited spatial resolution
of the Fermi-LAT data, it is difficult to draw a definite conclusion about
the nature of the residual emission yet. Later we will show that, such
emission would even degenerate with that from some DM annihilation models.
However, as will be discussed in Sec. 4, such emission is less likely to 
be of DM origin. Therefore we assume the IRIS + point model as the 
astrophysical background in the following analysis.

\begin{table}[htb]
\centering
\begin{tabular}{lcc}
\hline
\hline
Template & $-\ln{\mathcal L}$ & $\Delta N_{\rm dof}$ \\
\hline
IRIS & 190463.0 & --- \\
IRIS + point & 190454.9 & 4 \\
$0^{\circ}.9$ disk$^a$ & 190456.1 & 1 \\
IRIS + $0^{\circ}.9$ disk$^a$ & 190455.7 & 3 \\
IRIS + two bubbles$^a$ & 190453.8 & 3 \\
\hline
\hline
\end{tabular}\\
Note: $^a$The radius of the disk or bubbles contributes 1 to the number
of degree-of-freedom.
\caption{\label{tab:fit}Comparison of the goodness-of-fit of various
spatial templates.}
\end{table}

\begin{figure}
\centering
\includegraphics[width=.6\textwidth]{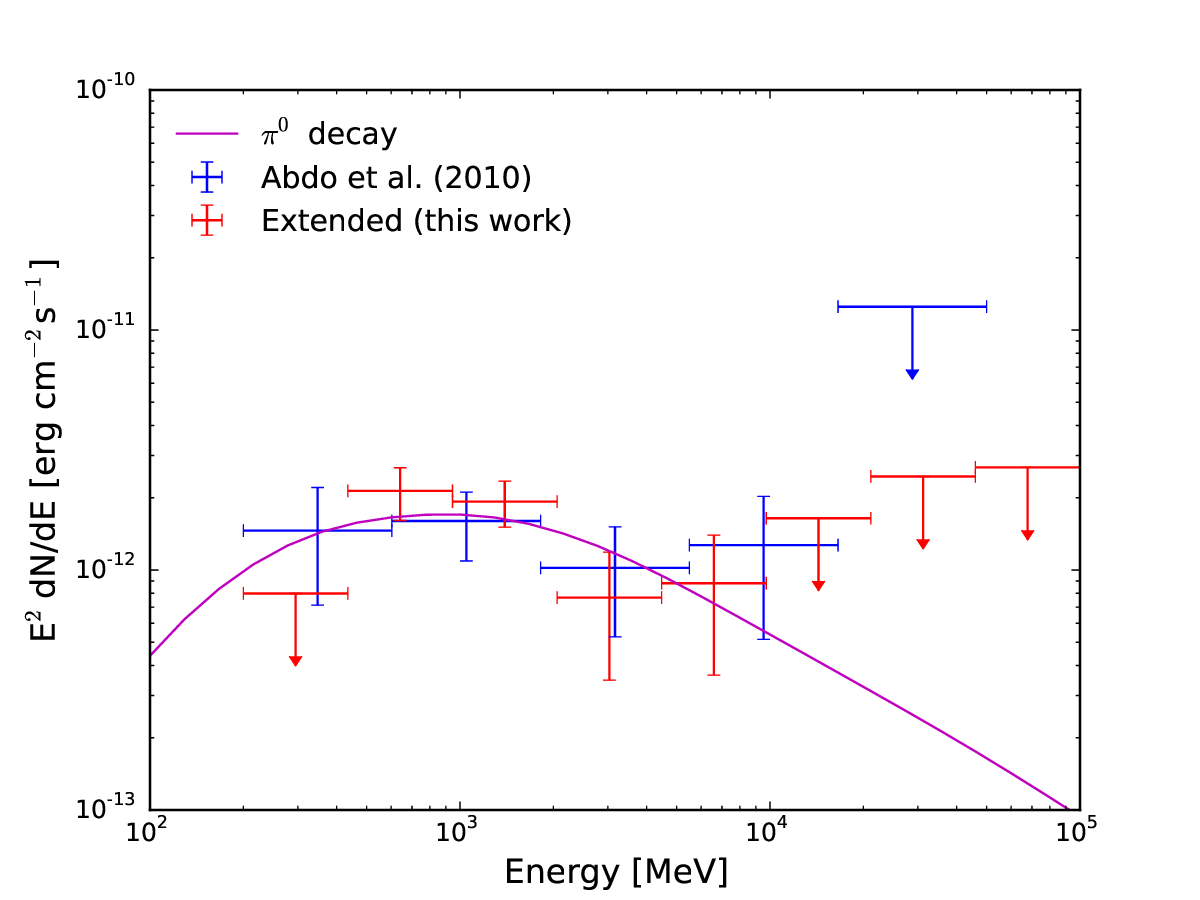}
\caption{SED of M31 obtained in this work (red), compared with that reported 
in Ref. \cite{2010A&AAbdo}. The solid line shows the expectation from the
$pp$-collision-induced $\gamma$-ray spectrum for a proton spectrum similar 
to that of the Milky Way.}
\label{fig:sed}
\end{figure}

To derive the spectral energy distribution (SED) of M31, we divide the
data into eight energy bins from 200 MeV to 100 GeV, and use {\tt gtlike} 
to fit the flux of M31 (the IRIS component) in each bin. During the fit
we fix the spectral parameters of all sources to the values derived in 
the global fitting, and leave the normalizations of all point sources 
within the ROI, and the normalizations of the diffuse backgrounds free. 
The results are shown in Figure \ref{fig:sed} (red). Also shown are the 
SED obtained in Ref. \cite{2010A&AAbdo} based on two year data (blue). 
These two results are consistent with each other within statistical 
errors. We find that the data look similar to that from the decay of 
neutral pions produced by the interaction between cosmic ray protons
and gas, as shown by the solid line. Here we parameterize the proton
spectrum as $dN/dE_k\propto\left(1+E_k/1.6\,{\rm GeV}\right)^{-2.8}$, 
which is similar to that of the Milky Way.

\section{Constraints on DM annihilation models}

\subsection{Prompt radiation only}

Now we discuss the potential emission from the annihilation of DM. In this 
subsection we consider only the prompt radiation produced directly 
associated with the DM annihilation. We add an extended source with spatial 
distribution proportional to the $J$-factor as expected from the DM 
annihilation and fit to the data. During the fitting we fix the spectral 
parameters of all sources in the ROI to the values obtained in the global 
fitting described in Sec. 3, and leave the normalizations of these sources 
and the diffuse backgrounds, as well as the normalization of the DM component 
free. The energy spectrum of the DM component is calculated by Pythia,
given the mass of the DM particle and the annihilation channel.

\begin{figure}
\centering
\includegraphics[width=.45\textwidth]{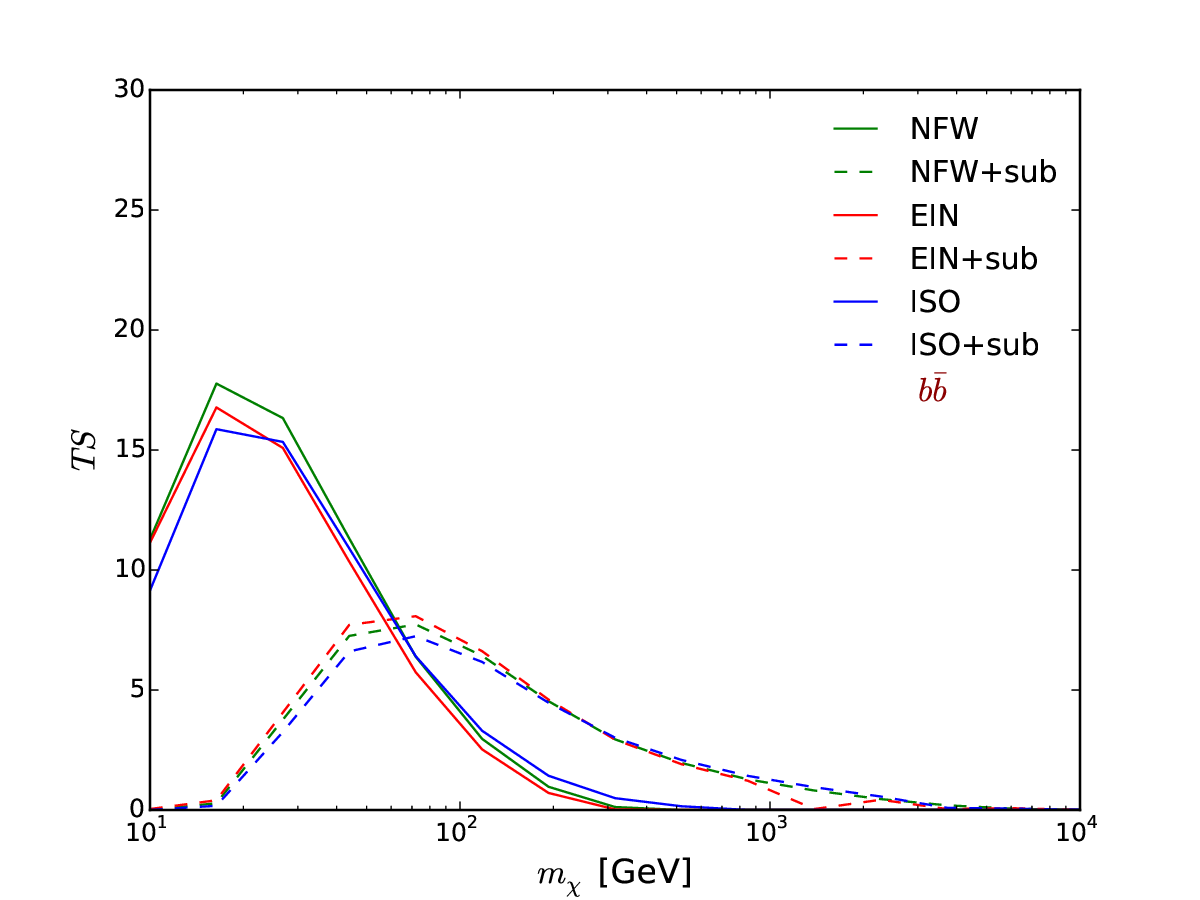}
\includegraphics[width=.45\textwidth]{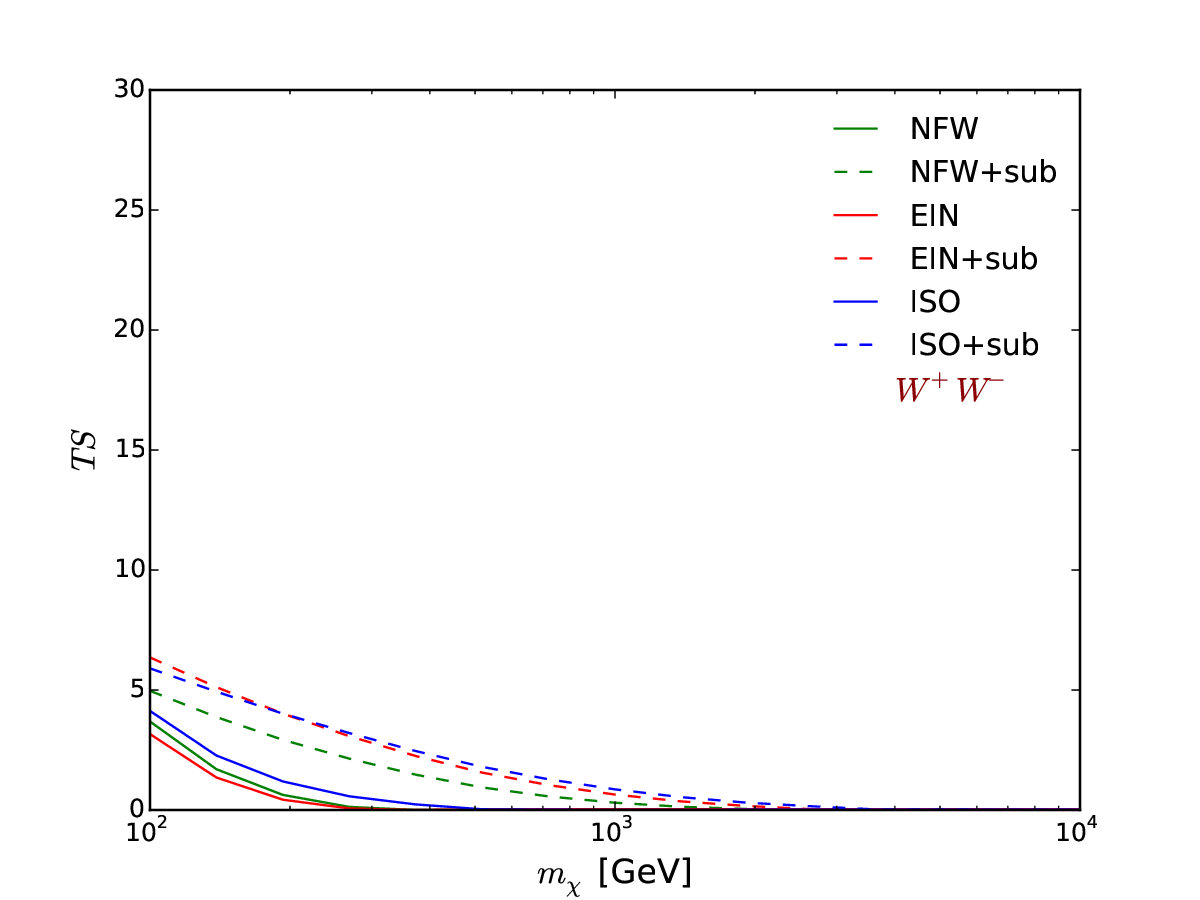}
\includegraphics[width=.45\textwidth]{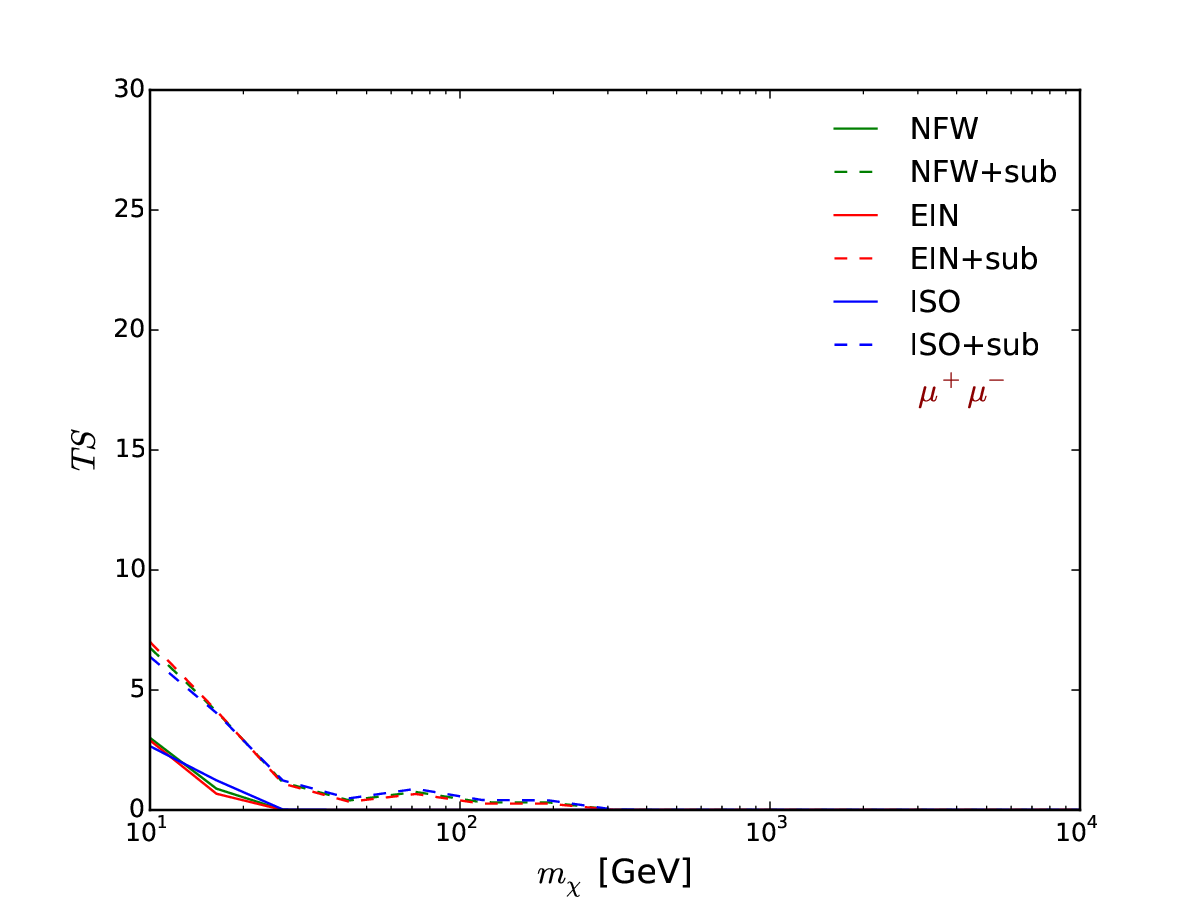}
\includegraphics[width=.45\textwidth]{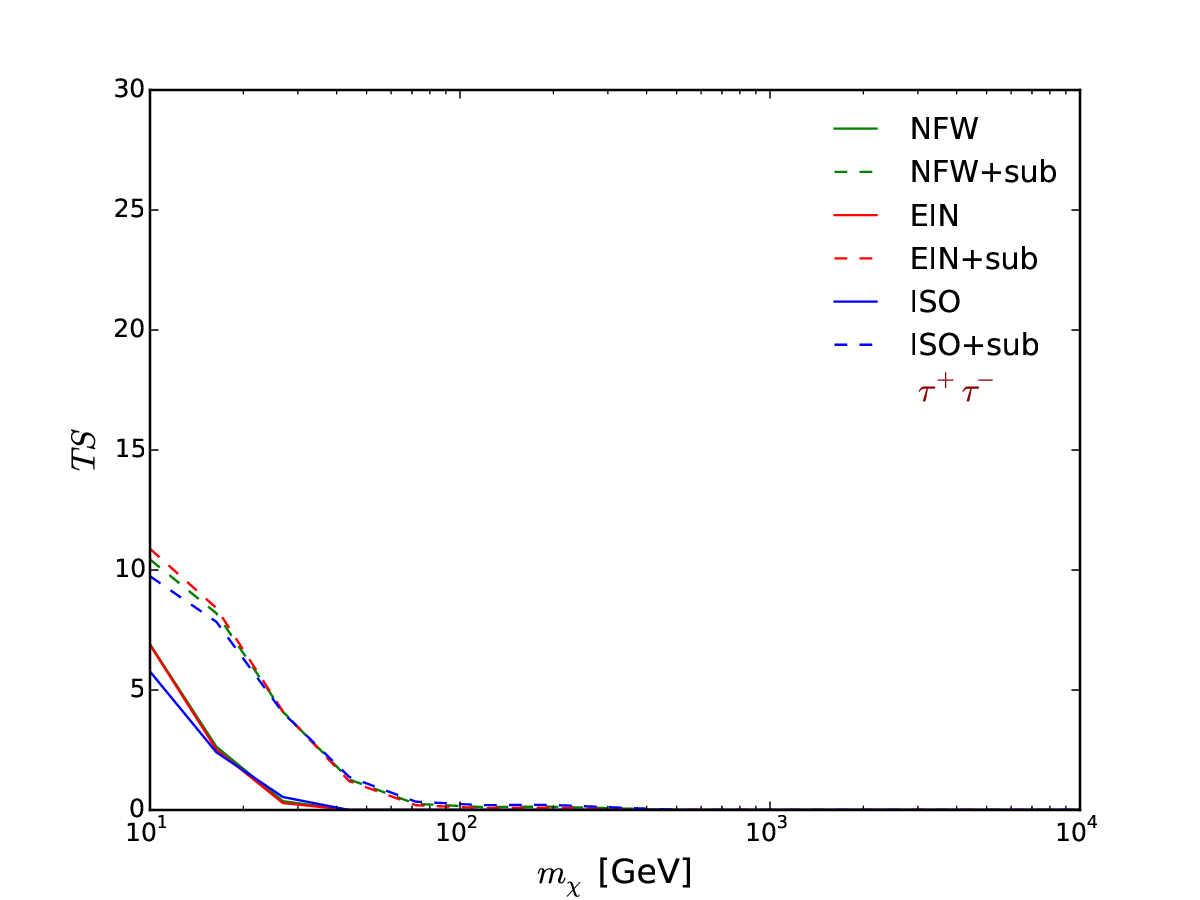}
\caption{TS values of the DM component for four annihilation channels:
$b\bar{b}$ (top-left), $W^{+}W^{-}$ (top-right), $\mu^+\mu^-$ (bottom-left), 
and $\tau^+\tau^-$ (bottom-right).}
\label{fig:TS}
\end{figure}

We find that in some cases the DM annihilation component degenerates
with the astrophysical background emission of M31. In particular, for 
$m_{\chi}\sim 16$ GeV and $b\bar{b}$ annihilation channel, the inclusion 
of the DM annihilation component from the main halo gives $-\ln{\mathcal L}
=190446$. It corresponds to a TS value of the DM component of $\sim 18$
compared with the {\it null} hypothesis ($-\ln{\mathcal L}_0=190454.9$), 
as shown in Figure \ref{fig:TS}\footnote{For $W^+W^-$, $\mu^+\mu^-$, and 
$\tau^+\tau^-$ channels the TS values are smaller, because the $\gamma$-ray 
spectra from these final states are harder and deviate more from the 
background.}. In this case the DM component also shares a large fraction 
of the flux of M31, resulting in a very low significance of M31 itself. 
However, it is unlikely that the data favor a DM component because the 
inferred cross section of $\sim10^{-25}-10^{-24}$ cm$^3$s$^{-1}$ (see 
below Figure \ref{fig:cross}) is orders of magnitude higher than the 
constraints from the Fermi-LAT observations of dwarf spheroidal galaxies 
\cite{Ackermann:2015zua}. When DM subhalos are taken into account, the 
degeneracy is broken and the corresponding TS value of the DM component 
decreases significantly, which implies that the morphology of the 
$\gamma$-ray emission is not traced by the $J_{\rm tot}$ distribution. 
Therefore we assume that the excess emission from the direction of M31 
has an astrophysical origin, and place upper limits on the DM model 
parameters instead. The degeneracy between the DM signal and the 
astrophysical background, however, makes these upper limits be higher, 
and the corresponding constraints on the DM cross section be more 
conservative.

\begin{figure}
\centering
\includegraphics[width=.45\textwidth]{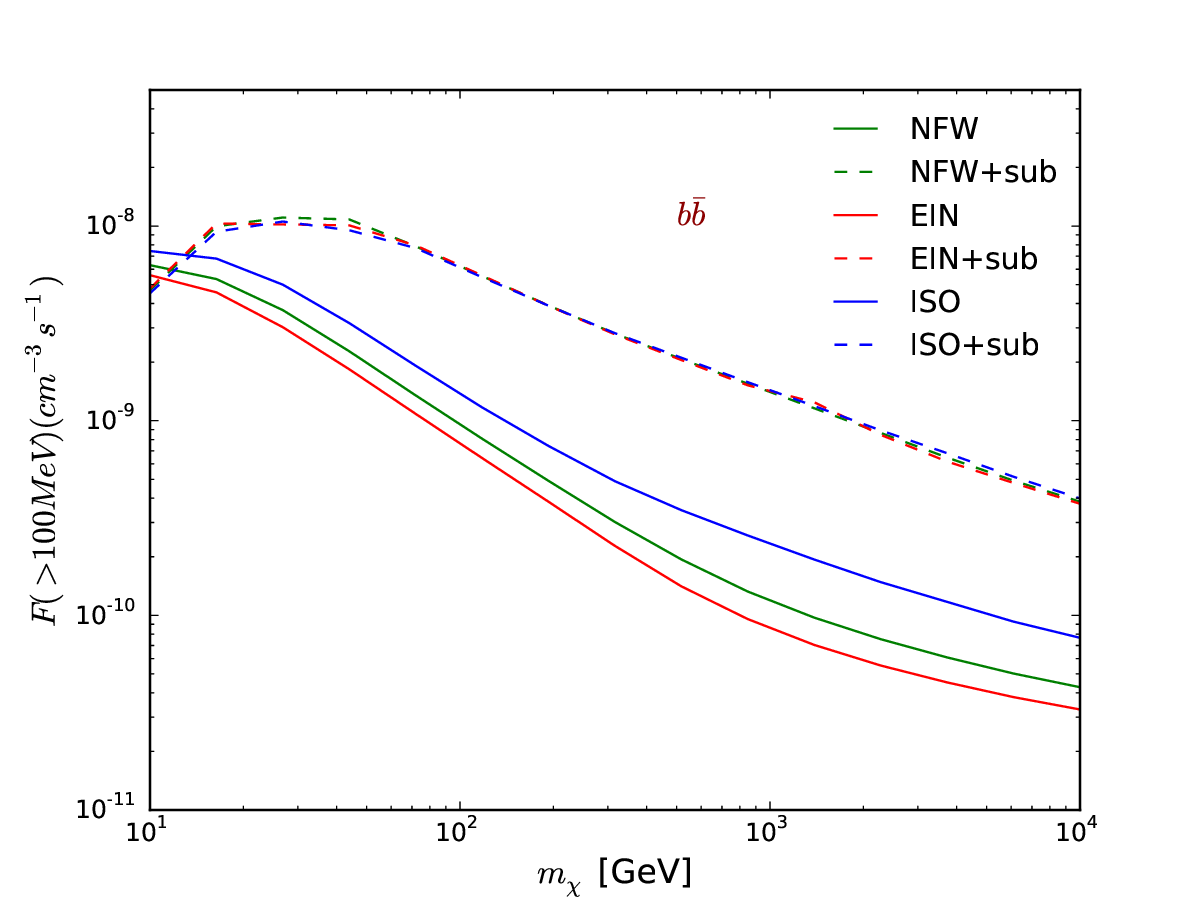}
\includegraphics[width=.45\textwidth]{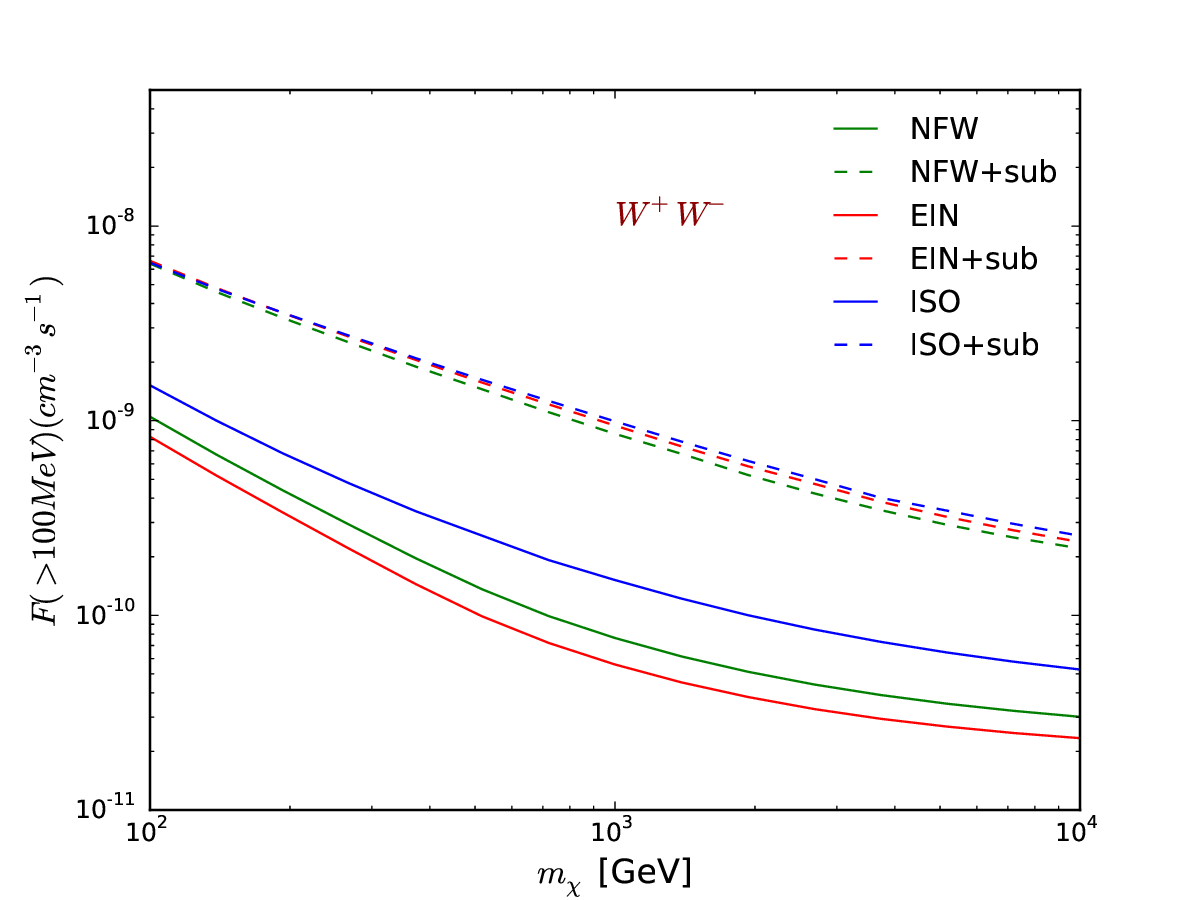}
\includegraphics[width=.45\textwidth]{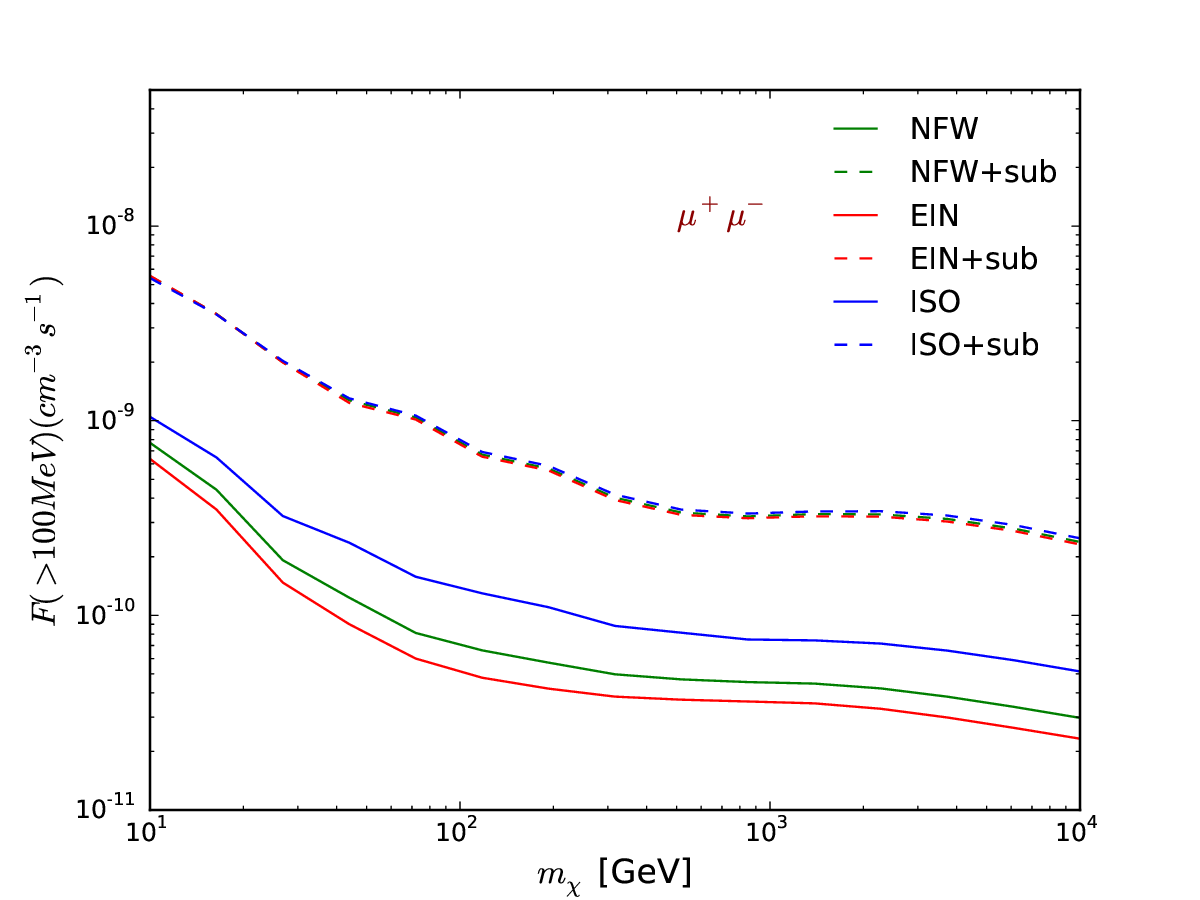}
\includegraphics[width=.45\textwidth]{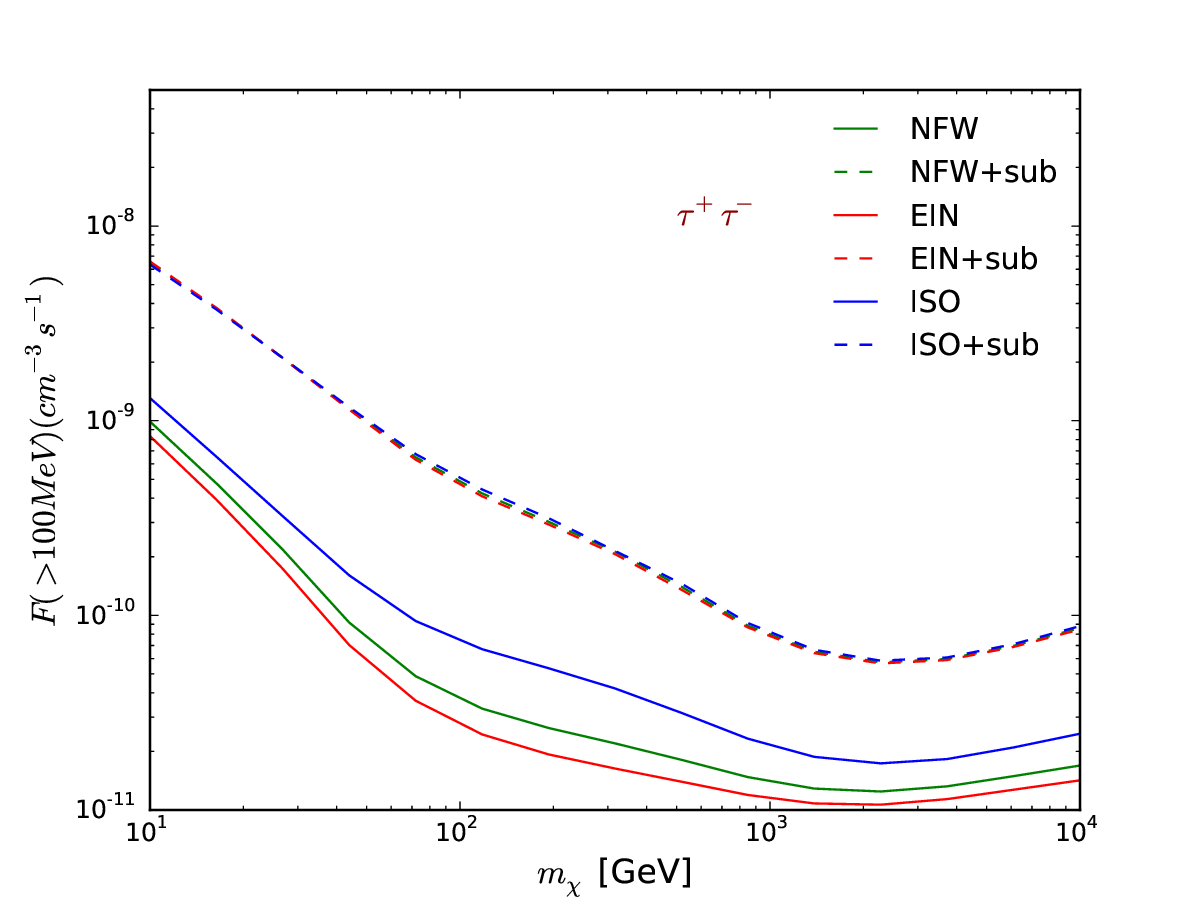}
\caption{$95\%$ upper limits of $>100$ MeV $\gamma$-ray flux of the
DM annihilation component, for $b\bar{b}$ (top-left), $W^{+}W^{-}$
(top-right), $\mu^+\mu^-$ (bottom-left), and $\tau^+\tau^-$ 
(bottom-right) channels, respectively. See the text for details.}
\label{fig:flux}
\end{figure}

The upper limits for the $\gamma$-ray fluxes above 100 MeV from DM 
annihilation of M31 for $b\bar{b}$, $W^+W^-$, $\mu^+\mu^-$, and
$\tau^+\tau^-$ channels are shown in Figure \ref{fig:flux}. The lower 
(higher) group in each panel represents the cases without (with) subhalos.
If subhalos are not taken into account, the derived upper limits differ 
from each other for different assumed density profiles. The flatter the 
density profile, the higher the flux limit. When subhalos are included, 
we find that the flux upper limits are quite similar for different profiles 
of the smooth halo. This is because the contribution from subhalos dominate
over the main halo, and we assume the same subhalo distribution (Eq. 
(\ref{eq:sub})) in this work. However, since the $J$-factors will be 
different for the three main halo profiles, we will get different
constraints on the DM model parameters (see below).

\begin{figure}
\centering
\includegraphics[width=.45\textwidth]{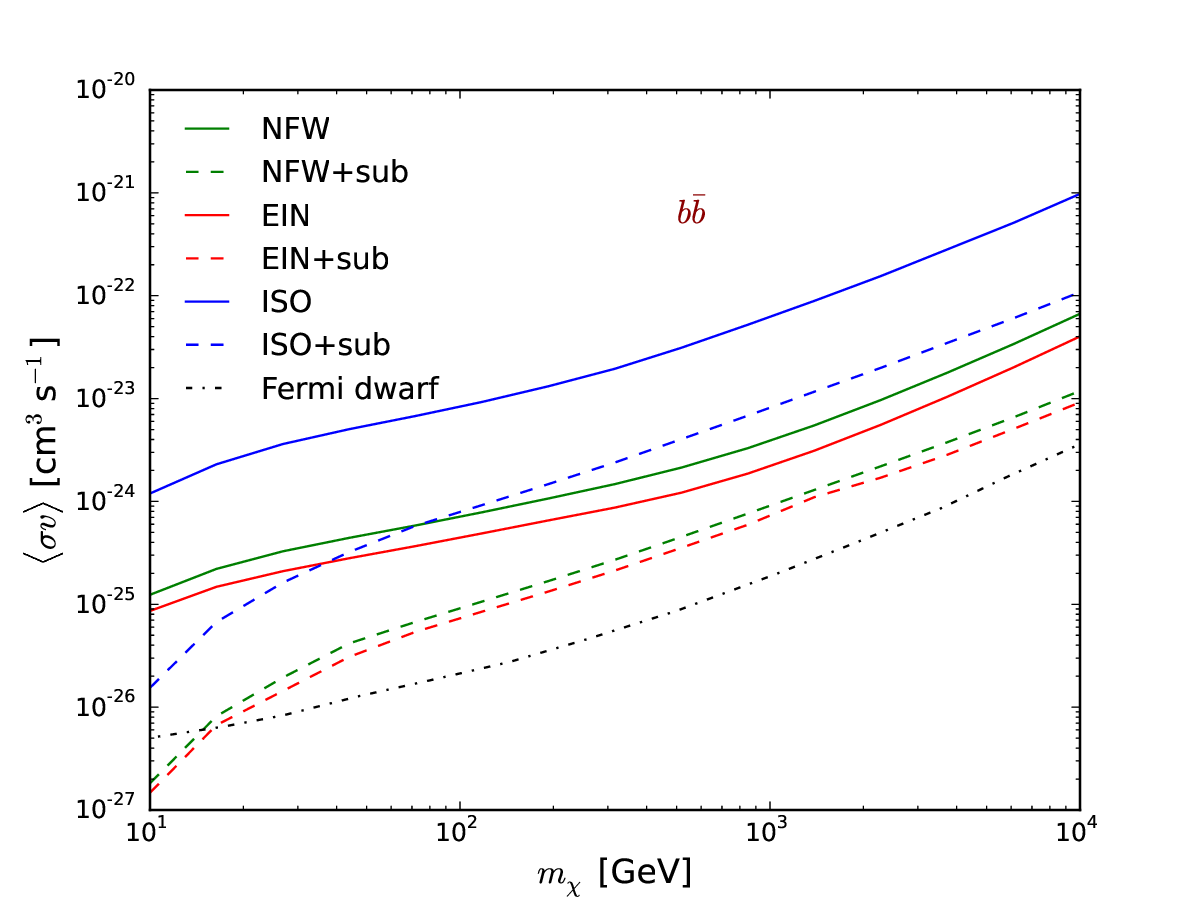}
\includegraphics[width=.45\textwidth]{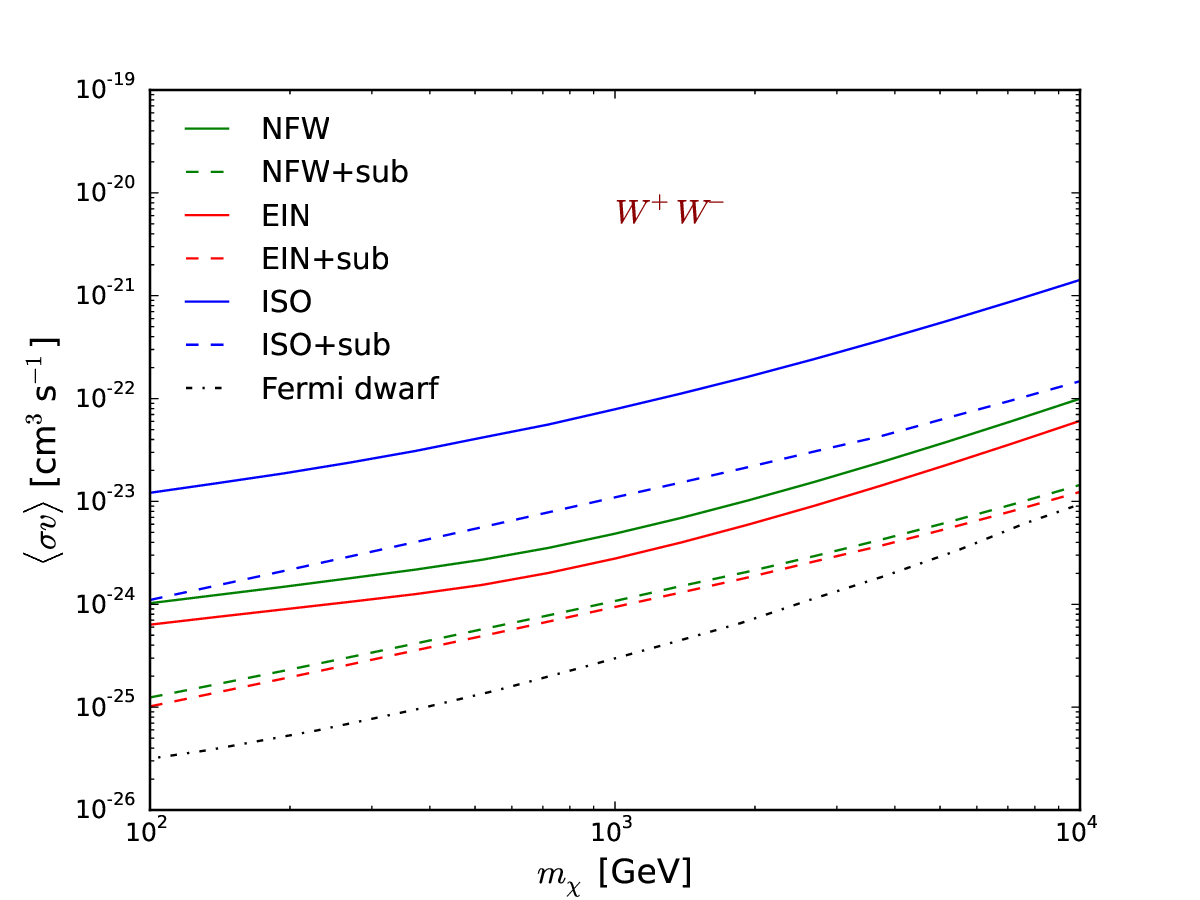}
\includegraphics[width=.45\textwidth]{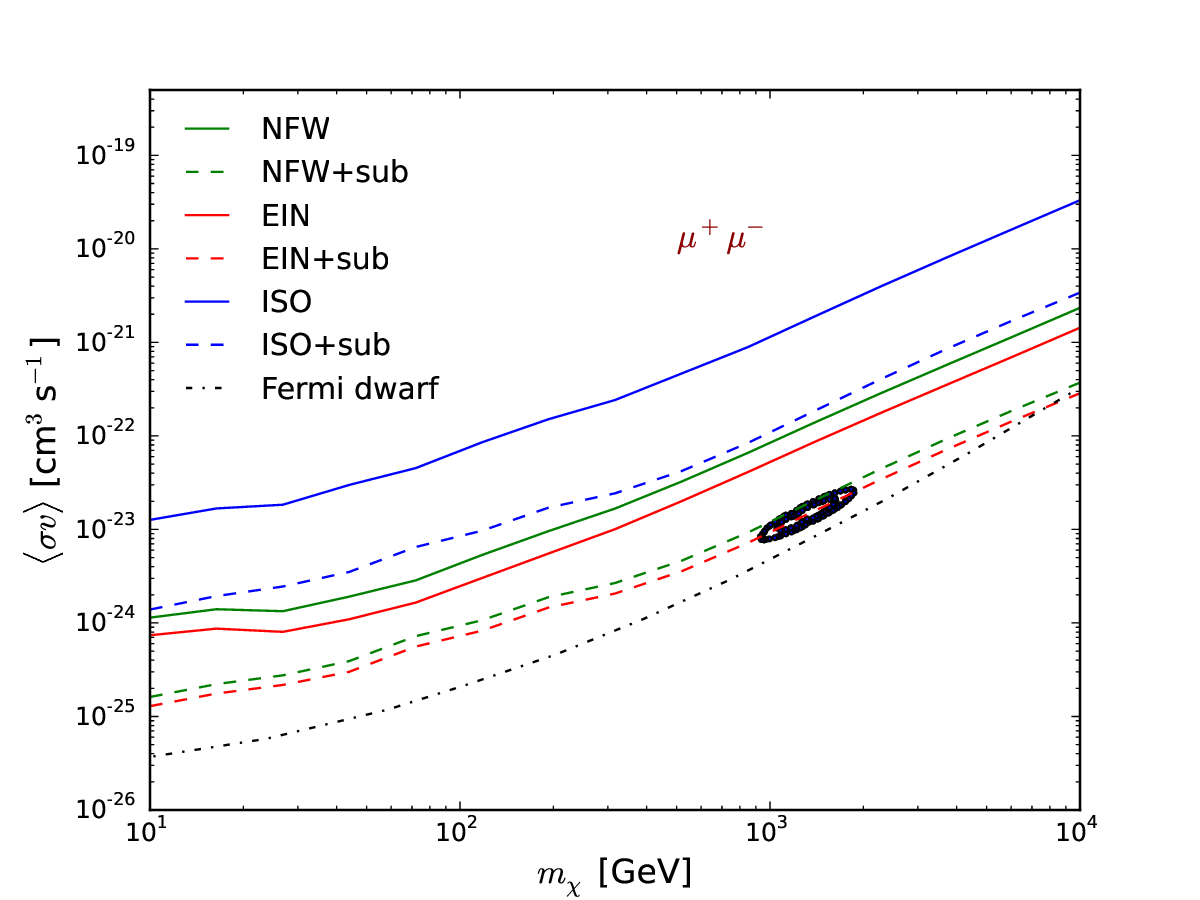}
\includegraphics[width=.45\textwidth]{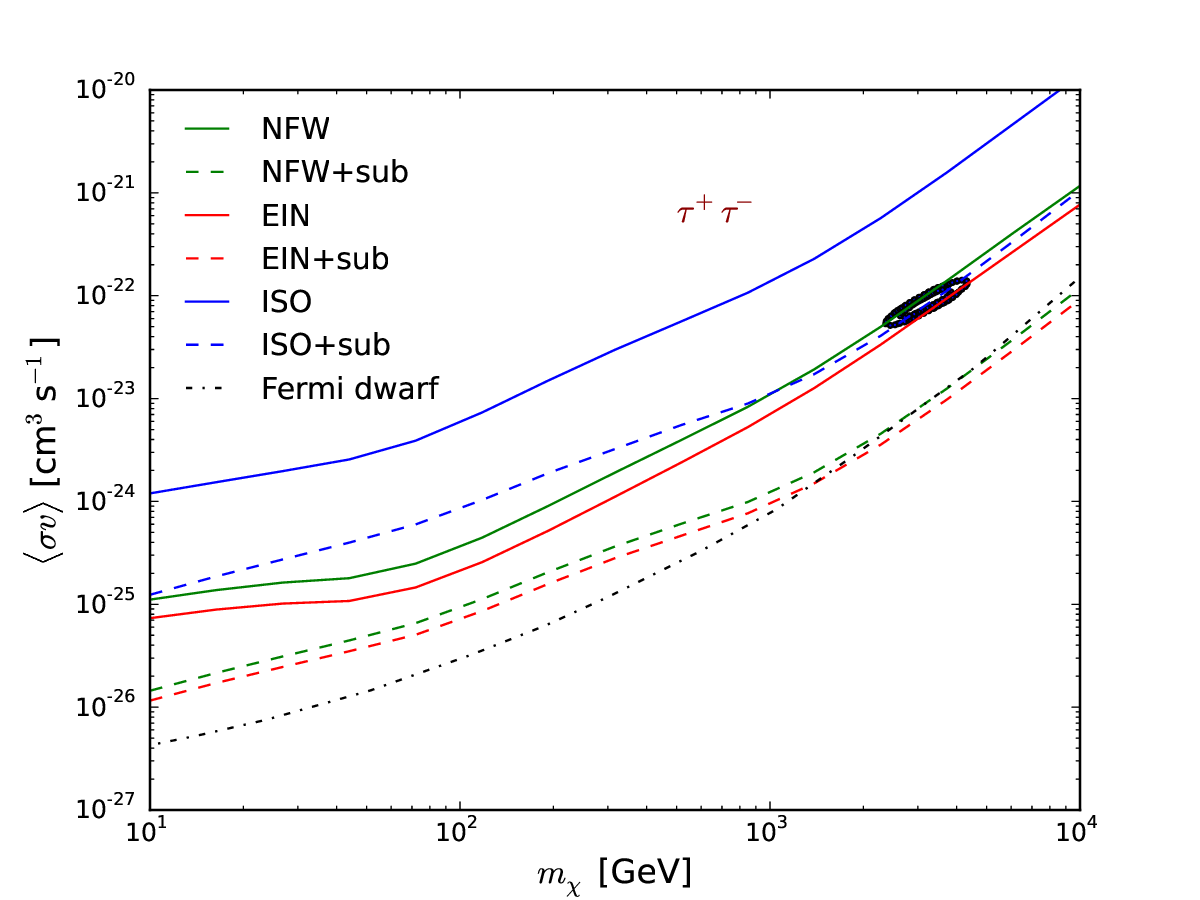}
\caption{$95\%$ confidence level constraints on DM annihilation cross 
section for the four annihilation channels as shown in Figure \ref{fig:flux}.
For comparison, the black dot-dashed line shows the result from the
combined analysis of Fermi-LAT data of 15 dwarf spheroidal galaxies
\cite{Ackermann:2015zua}.}
\label{fig:cross}
\end{figure}

The 95\% upper limits on the DM annihilation cross sections are presented 
in Figure \ref{fig:cross}. For comparison we also show the results from the 
combined analysis of Fermi-LAT observations of 15 dwarf spheroidal galaxies
\cite{Ackermann:2015zua}. It is shown that the constraints for NFW and EIN 
profiles are comparable with each other, while for ISO profile they are 
weaker by almost an order of magnitude. Including subhalos will improve 
the limits by a factor of $\sim10$, although the flux limits are higher.
Compared with the results from dwarf galaxies, the constraints from M31
will be generally weaker if the subhalos are not considered. If subhalos
are taken into account and the density profile of the main halo is cuspy, 
our derived constraints are comparable to (or slightly weaker than) that
from the population of dwarf spheroidal galaxies.

The lepton pair channels $\mu^+\mu^-$ and $\tau^+\tau^-$ are motivated
by the recent observations of electron/positron excesses
\cite{2008NaturChang,2009NaturAdriani,2009A&AAharonian,2009PhRvLAbdo,
2013PhRvLAguilar}, and non-excess of antiprotons
\cite{2009PhRvLAdriani,2010PhRvLAdriani}. In the lower two panels of
Figure \ref{fig:cross} we show the required parameter regions to fit 
the electron/positron excesses measured by PAMELA/Fermi-LAT/AMS-02
\cite{2013PhLBYuan,Yuan:2014pka}. It is shown that for $\mu^+\mu^-$ 
channel the current limits from M31 can marginally constrain the $e^+e^-$ 
excess favored parameter regions. For DM annihilation into $\tau^+\tau^-$, 
the model will predict too many $\gamma$-ray photons and is in significant
conflict with the Fermi-LAT observations of M31.

\subsection{Including ICS emission}

In the above discussion the ICS contribution from DM annihilation induced
$e^+e^-$ is not included. Here we discuss the effect of the ICS component. 
We take NFW profile and $\mu^+\mu^-$ annihilation channel as an example.
DM subhalos are not taken into account. A spherical geometry of the 
electron propagation in the halo of M31 is assumed. The propagation 
equation can be written as
\begin{equation}
\nabla\cdot\left[D(E,{\bf r})\nabla\frac{dn}{dE}\right]+\frac{\partial}
{\partial E}\left[b(E,{\bf r})\frac{dn}{dE}\right]+Q(E,{\bf r})=0,
\label{diff}
\end{equation}
where $dn/dE$ is the equilibrium density distribution of $e^+e^-$, 
$D(E,{\bf r})$ is the diffusion coefficient, $b(E,{\bf r})=-dE/dt$ is the 
absolute energy loss rate, and $Q(E,{\bf r})$ is the source injection rate.
We assume a homogeneous diffusion coefficient, $D(E)=3\times10^{28}
(E/{\rm GeV})^{1/3}$, as that in the Milky Way. Since here we consider a 
large halo of M31, we neglect the interstellar radiation field from stars 
and dust, and only consider the cosmic microwave background (CMB) photons 
to calculate the cooling and ICS emission of the $e^+e^-$. The cooling 
rate is then $b(E)=2.5\times10^{-17}(E/{\rm GeV})^2$ GeV s$^{-1}$. 

The solution of Eq. (\ref{diff}) is \cite{2006A&AColafrancesco}
\begin{equation}
\frac{dn}{dE}=\frac{1}{b(E)}\int_E^{\infty}dE'G(r,\Delta v)Q(E',r),
\end{equation}
where
\begin{eqnarray}
G(r,\Delta v)&=&\frac{1}{\sqrt{4\pi\Delta v}}\sum_{n=-\infty}^{+\infty}
\int_0^{r_h}dr'\frac{r'}{r_n}\nonumber\\
&\times&\left[\exp\left(-\frac{(r'-r_n)^2}{4\Delta v}\right)-\exp\left(
-\frac{(r'+r_n)^2}{4\Delta v}\right)\right]\frac{\rho^2(r')}{\rho^2(r)},
\end{eqnarray}
$r_h=r_{\rm vir}$, $r_n=(-1)^nr+2nr_h$, and $\Delta v(E,E')=\int_E^{E'}
de\,D(e)/b(e)$. Given the equilibrium $e^+e^-$ spectrum, the ICS emission 
at each location $r$ can be calculated using the Klein-Nishina differential
scattering cross section \cite{Blumenthal:1970gc,2010ApJZhang}.

For each mass of the DM particle, we calculate the $\gamma$-ray (both the 
ICS and the prompt radiation) emissivity as a function of photon energy 
$E$ and radius $r$, and integrate along the l.o.s. of given direction 
$\theta$. A skymap in each energy bin can be obtained. Similar to the 
Galactic diffuse $\gamma$-ray template, we build a template cube of the 
DM-induced photon emission which is embedded in the model for the fitting.

\begin{figure}
\centering
\includegraphics[width=.45\textwidth]{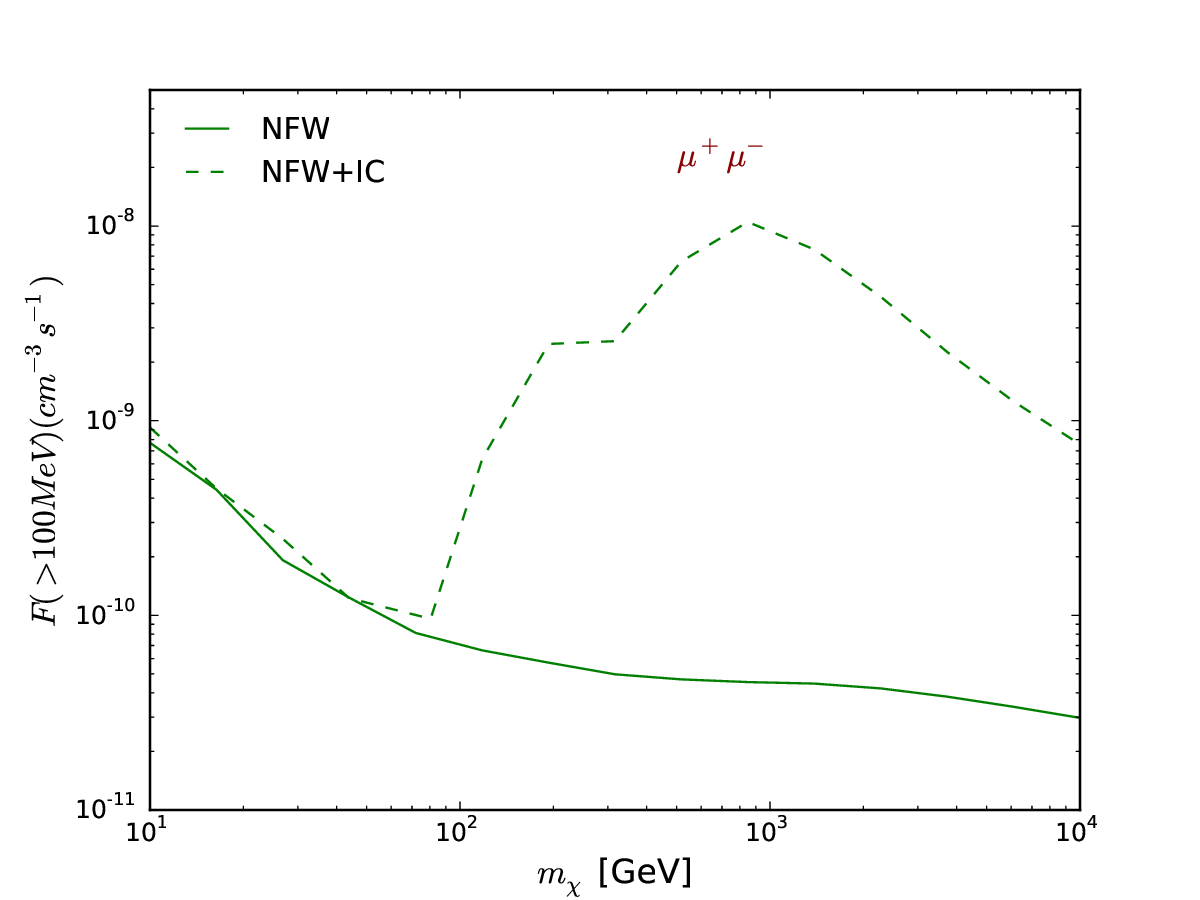}
\includegraphics[width=.45\textwidth]{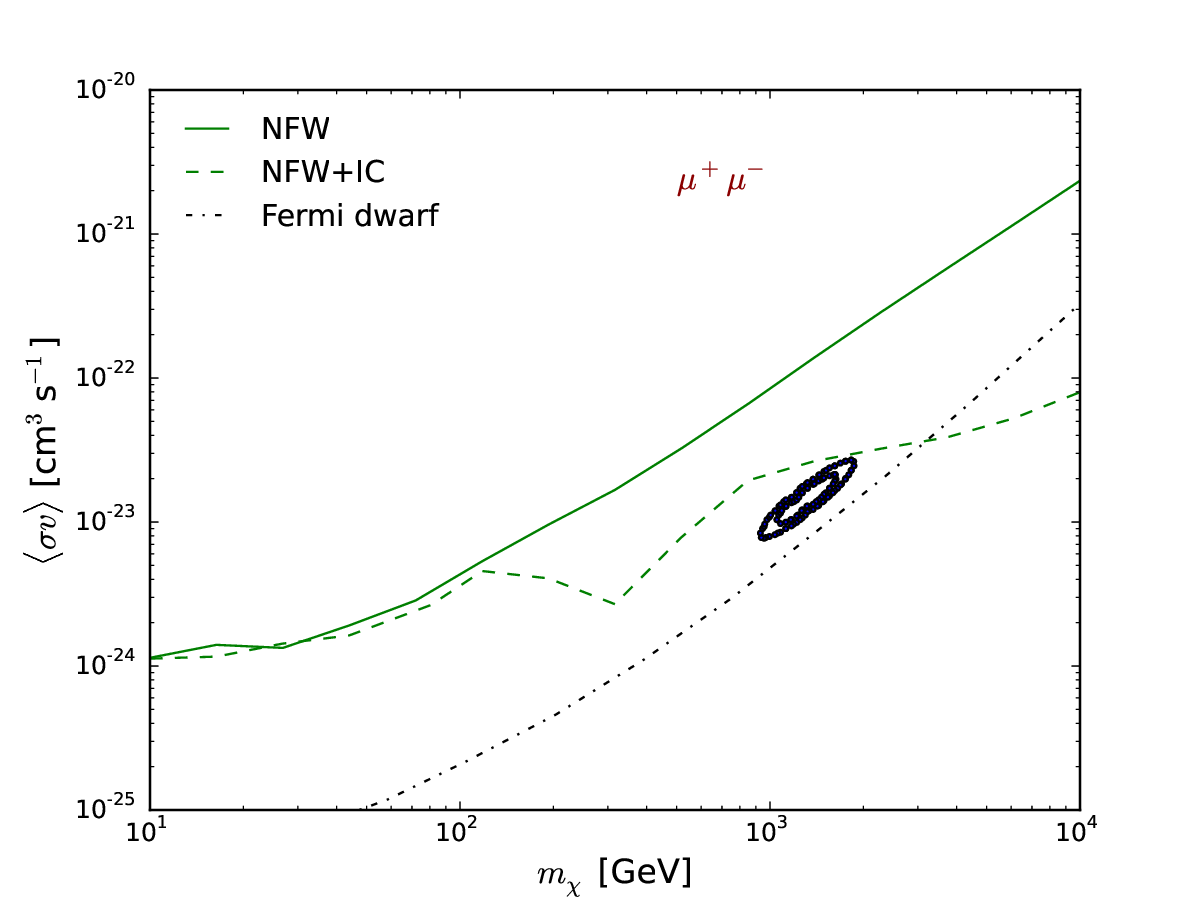}
\caption{$95\%$ confidence level upper limits of the $>100$ MeV $\gamma$-ray
flux (left) and the DM annihilation cross section (right) when the ICS 
component is taken into account. The annihilation channel is $\mu^{+}\mu^{-}$ 
and the density profile of the main halo is NFW.}
\label{fig:IC}
\end{figure}

The results are shown in Figure \ref{fig:IC}. The left panel shows the
flux upper limits, and the right panel shows the constraints on the DM
cross section. For $m_{\chi}\lesssim100$ GeV the ICS photon energies
are essentially smaller than $100$ MeV and the Fermi-LAT data cannot give
effective constraint. For $m_{\chi}>100$ GeV the ICS component enters
the Fermi-LAT energy region. The flux upper limits show a remarkable
increase for $m_{\chi}>100$ GeV, possibly because, compared with the 
prompt radiation, the spatial distribution of the ICS emission is more 
extended and the spectrum is softer, which mimics the background more.
Nevertheless, the constraints on $\langle\sigma v\rangle$ are stronger 
than the case with only the prompt radiation, as shown in the right panel 
of Figure \ref{fig:IC}. For $m_{\chi}\gtrsim$TeV the constraints can 
improve by a factor of a few to $\sim10$. 

\section{Conclusion and discussion}

In this work, we analyze 7.5 year Fermi-LAT $\gamma$-ray data of M31 to 
probe the particle DM annihilation models. We confirm the results in Ref. 
\cite{2016MNRAS.459L..76P} that there is residual excess from the vicinity 
of M31 besides the extended emission from the galactic disk as traced by 
the far infrared dust emission. However, the detailed morphology of the 
residual is not clear yet, which needs further studies with improved
photon statistics and angular resolution. When adding an additional point 
source located at $\sim0^{\circ}.9$ away from the galaxy center, a 
relatively clean residual TS map is obtained. Slight improvements of
the fittings are found if an extended uniform disk or two bubble-like 
templates are added instead of the point source. These results may
imply that the dust emission template is not good enough to describe
the $\gamma$-ray emission from M31. As we learn from the Milky Way,
the diffuse $\gamma$-ray emission should be the convolution of the cosmic 
ray distribution and the gas distribution. Furthermore, the ICS emission 
from cosmic ray electrons will also make the $\gamma$-ray distribution 
deviate from the dust distribution. Finally, a few bright point sources
such as pulsars may contribute effectively to the $\gamma$-ray emission 
of an extragalactic galaxy \cite{Ackermann:2015whl}.

Using the far infrared dust template and an additional point source as
the background model of M31, we search for the DM annihilation signals.
We find that for $m_{\chi}\sim$ a few tens of GeV and $b\bar{b}$
annihilation channel, the DM emission from the main halo (i.e., without
subhalos) degenerates with the galaxy emission significantly. If the
DM annihilation from a smooth NFW/EIN halo is employed to explain the
majority of the $\gamma$-ray emission, the required cross section is
$\sim10^{-25}$ cm$^3$ s$^{-1}$, which is well above the exclusion limits
from dwarf galaxy observations \cite{Ackermann:2015zua}. Therefore it is 
unlikely that DM annihilation dominates the $\gamma$-ray emission of M31. 

If the subhalos are taken into account, the degeneracy is broken and more 
stringent constraints on the DM parameter space can be derived. For NFW 
and EIN profiles, the constraints are stronger by a factor of $\sim10$ 
than that for ISO profile. The most stringent constraints (for NFW or EIN 
profile with subhalos) set by the Fermi-LAT observations of M31 are 
comparable to (for leptonic channels) or slightly weaker than (for quark and
gauge boson channels) that of the combined analysis of 6 year observations 
of 15 dwarf galaxies \cite{Ackermann:2015zua}. The parameter regions to 
explain the $e^+e^-$ excesses will be excluded if the DM annihilation 
luminosity is enhanced by subhalos.

We also test the effect of including the ICS component from DM annihilation
induced electrons/positrons. For the first time we employ an energy dependent 
template cube of the ICS emission in the data analysis to compute the
likelihood and upper limits. The resulting constraints on $\langle\sigma 
v\rangle$ are improved remarkably for $m_{\chi}\gtrsim 100$ GeV.

\acknowledgments
We thank the anonymous referee for helpful suggestions, and Eric Albin, 
Yi-Zhong Fan, Zhao-Qiang Shen, and Zi-Qing Xia for useful discussion. 
Q.Y. is supported by the National Key Program for Research and Development 
(No. 2016YFA0400200) and the 100 Talents program of Chinese Academy of
Sciences.

\bibliographystyle{JHEP}
\bibliography{refs}
\end{document}